\UseRawInputEncoding
\documentclass[12pt,technote, onecolumn, draft]{IEEEtran}

\usepackage{latexsym}
\usepackage{mathrsfs}
\usepackage{multirow}
\usepackage{amsfonts,amssymb,amsmath,amsthm,bm}
\usepackage{color}
\usepackage{cite}
\usepackage{lmodern} 

\usepackage{graphics,arydshln}
\usepackage{autobreak}
\allowdisplaybreaks

\newtheorem{theorem}{Theorem}
\newtheorem{example}{Example}
\newtheorem{definition}{Definition}
\newtheorem{remark}{Remark}
\newtheorem{corollary}{Corollary}
\newtheorem{lemma}{Lemma}
\newtheorem{proposition}{Proposition}

\def\BF{\mathbb{F}}
\def\bw{\mathbf{w}}
\def\bv{\mathbf{v}}
\def\bu{\mathbf{u}}
\def\bH{\mathbf{H}}
\def\bO{\mathbf{O}}
\def\bI{\mathbf{I}}
\def\bA{\mathbf{A}}
\def\bP{\mathbf{P}}
\def\bQ{\mathbf{Q}}
\def\bM{\mathbf{M}}
\def\bG{\mathbf{G}}
\def\bF{\mathbf{F}}
\def\bE{\mathbf{E}}
\def\bN{\mathbf{N}}
\def\bX{\mathbf{X}}
\def\bB{\mathbf{B}}

\def\bW{\mathbf{W}}
\def\bU{\mathbf{U}}
\def\bD{\mathbf{D}}

\def\hcC{\hat{\mathcal{C}}}

\def\cF{\mathcal{F}}
\def\cD{\mathcal{D}}
\def\cS{\mathcal{S}}
\def\cC{\mathcal{C}}
\def\cA{\mathcal{A}}
\def\cH{\mathcal{H}}
\def\cP{\mathcal{P}}
\def\cQ{\mathcal{Q}}
\def\cX{\mathcal{X}}

\def\cM{\mathcal{M}}
\def\cB{\mathcal{B}}

\def\cN{\mathcal{N}}
\def\cX{\mathcal{X}}

\def\cF{\mathcal{F}}
\def\cG{\mathcal{G}}

\def\hcB{\widehat{\mathcal{B}}}
\def\hcN{\widehat{\mathcal{N}}}
\def\hcC{\widehat{\mathcal{C}}}

\def\hbB{\widehat{\mathbf{B}}}
\def\hbE{\widehat{\mathbf{E}}}
\def\hbN{\widehat{\mathbf{N}}}
\def\hbG{\widehat{\mathbf{G}}}

\def\hU{\widehat{U}}
\def\tcC{\Tilde{\mathcal{C}}}
\def\tcF{\Tilde{\mathcal{F}}}
\def\tbv{\Tilde{\mathbf{v}}}

\def\sL{\mathscr{L}}
\def\rank{\mathrm{rank}}
\def\rs{\mathrm{rs}}
\def\wt{\mathrm{wt}}
\def\RREF{\mathrm{RREF}}

\begin{document}

\title{New Constant Dimension Codes From the Inserting Mixed Dimension Construction and Multilevel Construction$^\dag$}
\author{Han Li, Fang-Wei Fu
\IEEEcompsocitemizethanks{\IEEEcompsocthanksitem Han Li and Fang-Wei Fu are with Chern Institute of Mathematics and LPMC, Nankai University, Tianjin 300071, China (Emails: hli@mail.nankai.edu.cn, fwfu@nankai.edu.cn).}

\thanks{$^\dag$This research is supported by the National Key Research and Development Program of China (Grant No. 2022YFA1005000), the National Natural Science Foundation of China (Grant Nos. 12141108 and 62371259), the Fundamental Research Funds for the Central Universities of China (Nankai University), and the Nankai Zhide Foundation.}
}

\maketitle

\begin{abstract}
Constant dimension codes (CDCs) are essential for error correction in random network coding. A fundamental problem of CDCs is to determine their maximal
possible size for given parameters.  Inserting construction and multilevel construction are two effective techniques for constructing CDCs. We first provide a sufficient condition for a subspace to be added to the code from the mixed dimension construction in Lao et al. (IEEE Trans. Inf. Theory 69(7): 4333-4344, 2023). By appropriately combining matrix blocks  from small CDCs and rank-metric codes, we introduce three inserting constructions based on the mixed dimension construction. Furthermore, the mixed dimension construction and these inserting constructions 
are improved by the multilevel construction that is based on lifting rank-restricted Ferrers diagram rank-metric codes. Our constructions yield some new lower bounds for CDCs,  which are superior to the previously best-known ones.
\end{abstract}


\section{Introduction}
Subspace codes, particularly constant dimension codes, have garnered increasing attention since the seminal paper \cite{Network} of K\"{o}tter and Kschischang, in which they introduced a practical application of such codes in random network coding.

Let $\BF_q$ be the finite field of order $q$, and $\BF_q^n$ be the $n$-dimensional vector space over $\BF_q$. 
The projective space of order $n$ over $\BF_{q}$, denoted as $\cP_q(n)$, is the set of all $\BF_q$-subspaces of $\BF_q^n$.  Furthermore, for a non-negative integer $k \leq n$, the set of all $k$-dimensional subspaces of $\BF_q^n$ is called the Grassmannian $\cG_q(n,k)$ and its cardinality is given by the $q$-ary Gaussian binomial coefficient \[\begin{bmatrix}n\\k\end{bmatrix}_{q}= \prod \limits_{i = 0}^{k - 1}\frac{q^n - q^i}{q^k-q^i}.\]
The projective space $\mathcal{P}_q(n)$ and the                           Grassmannian $\mathcal{G}_{q}(n,k)$ are metric spaces \cite{Network}, endowed with the natural measure $d_{S}(U,V) \triangleq \dim(U + V) - \dim(U \cap V) $, where $\dim(\cdot)$ denotes the dimension of a vector space over $\mathbb{F}_{q}$. A non-empty subset $\mathcal{C} \subseteq \mathcal{P}_{q}(n)$ is termed a subspace code. The minimum distance of a subspace code $\mathcal{C}$, which  contains at least two codewords, is defined  as:
\[d_S(\cC) = \min\{d_S(U, V): U, V \in \cC, U \neq V\}.\] We say that the set $\cC \subseteq \cP_q(n)$ is an $(n,d)_q$ subspace code if the minimum  distance of $\mathcal{C}$ is at least $d$. The dimension distribution of $\cC$ is characterized by $\eta_0(\cC), \eta_1(\cC), \dots, \eta_n(\cC)$, where $\eta_i(\cC)$ is defined as $|\{U \in \cC: \dim(U) = i\}|$. Here $|\cdot|$ denotes the cardinality of a set. 
Since a code often has many values with $\eta_i(\cC)=0$, these values are usually omitted from the list. Consider a set
$T \subseteq \{0,1,\dots,n\}$, we refer to $\cC$ as an $(n, d, T)_q$ subspace code if $\eta_i(\cC) >0$ for $i \in T$ and $\eta_i(\cC)=0$ for $i \notin T$. This code is  called a mixed dimension code (MDC) when $|T|>1$, and a constant dimension code (CDC) when $|T|=1$. In addition, an $(n,d,T)_q$ subspace code with $M$ codewords is denoted as an $(n,M,d,T)_q$ subspace code.  Let $A_{q}(n,d,\{k\})$ denote the maximum number of codewords among all $(n,d,\{k\})_q$ CDCs. Hence, an $(n,d,\{k\})_{q}$ CDC with cardinality $A_q(n,d,\{k\})$ is considered to be optimal.  A main problem in the study of CDCs is to determine the value of  $A_{q}(n,d,\{k\})$ for $(n,d,\{k\})_q$ CDCs.

\subsection{Known Results}

Several bounds on $A_q(n,d,\{k\})$ have been derived in \cite{Network}, including the Hamming type upper bound, the Gilbert type lower bound, and the Singleton type upper bound. Wang, Xing, and Safavi-Naini\cite{authentication} established a connection between linear authentication codes and CDCs, and they provided the Wang-Xing-Safavi-Naini bound on CDCs. Xia and Fu \cite{Jonhson} derived two Johnson type bounds for CDCs,  and demonstrated that the Steiner structures $S[1,l,kl]_q$ are optimal CDCs, i.e.,  $A_q(kl,2l,\{l\}) = \frac{q^{kl}-1}{q^l-1}$. Honold, Kiermaier, and Kurz \cite{JohnMDC} extended the underlying idea of the Johnson type bound of CDCs to MDCs. Two bounds on the size of subspace codes, analogous to the classical Gilbert-Varshamov and linear-programming bounds, were provided in \cite{Programming}.    The homepage http://subspacecodes.uni-bayreuth.de/ lists the currently best-known lower and upper bounds of $(n,d)_q$ subspace codes and $(n,d,\{k\})_q$ CDCs for $4 \leq n \leq 19$ and $q \in \{2,3,4,5,7,8,9\}$\cite{Table}.

Rank-metric codes (RMCs) have been widely used in the construction of CDCs. Let $\mathbb{F}_{q}^{m \times n}$ be the matrix space, endowed with the rank metric $d_{R} (\mathbf{A},\mathbf{B})\triangleq \text{rank}(\mathbf{A} - \mathbf{B})$. An $(m \times n, d)_{q}$ RMC is a subset of  $\mathbb{F}_{q}^{m \times n}$ in which the rank distance between any two distinct codewords is at least $d$. Furthermore, a rank-metric code is said to be linear if it forms an $\mathbb{F}_{q}$-linear subspace of $\mathbb{F}_{q}^{m \times n}$, denoted as an $[m\times n, d]_{q}$ RMC. The Singleton bound for rank-metric codes\cite{DELSARTE, 1985Theory} states that the maximum possible size of an $(m \times n, d)_{q}$ RMC is $\Delta(m, n, d)_q := q^{\max\{m,n\}(\min \{m,n\} - d + 1)}$. A code achieving this bound is called a maximum rank distance (MRD) code. It is noteworthy that linear MRD codes exist for all feasible parameters\cite{DELSARTE, 1985Theory}.

Silva et al. \cite{Lifting} constructed a $(k+n,2d,\{k\})_{q}$ constant dimension code $\cC$ by lifting a $(k \times n, d)_{q}$ MRD code $\cM$, where $\cC = \{\mathrm{rowspace}(\bI_k \mid \bM): \mathbf{M} \in \cM\}$. 
However, this lifting construction had limitations in terms of generating CDCs with a large number of codewords. Later, the multilevel construction~\cite{Multilevel} was presented, which generalized the lifting construction and has inspired  a significant amount of research.  Utilizing the concept of  pending dots, Trautmann and Rosenthal\cite{penddot} enhanced the multilevel construction. Gluesing-Luerssen and Troha \cite{Linkage}
linked matrix representations of subspace codes of shorter length with the aid of a RMC, which resulted in a CDC of longer length without compromising the
distance. Xu and Chen \cite{Parallel} gave a construction of CDCs from two
parallel versions of lifted MRD codes. Since then, some new methods related to parallel linkage construction and multilevel construction have been proposed in \cite{GerLinkage}, \cite{ParMul},\cite{LinkMul}, \cite{ParLink}, \cite{MulLink}, \cite{ParLinkMul}. Two flexible
parameter-controlled inserting constructions of CDCs, including the direct inserting construction and the multilevel type inserting construction, were introduced in \cite{parcontroInsert}. Through
well-arranged combinations of the matrix blocks and small CDCs, improvements to the direct inserting construction were detailed in \cite{GenInsert}, \cite{BlockInsert1}, and \cite{BlockInser2}. The double multilevel construction, which combines the multilevel construction and the inverse multilevel construction, was presented in \cite{DoubleMul}, and can be seen as a generalization of the parallel multilevel construction \cite{ParMul}. Yu et al. \cite{2022Bilateral} extended the multilevel construction and the inverse multilevel construction to the bilateral multilevel construction. Subsequently, the generalized bilateral multilevel construction was proposed in \cite{GBilateral}. Some connections between CDCs and MDCs have been established in \cite{opt-F3}, \cite{Programming}, \cite{MDCconsb}, and \cite{MDC1}. Very recently, Lao et al. \cite{MixDD} introduced an interplay between constant dimension and mixed dimension codes, demonstrating that MDCs can be utilized to construct large CDCs. He et al. \cite{ParllelMixed} applied  the parallel construction to this mixed dimension construction, and obtained some new lower bounds on CDCs. 

\subsection{Our Contributions}
This paper is devoted to constructing CDCs with larger size.

Section \uppercase\expandafter{\romannumeral3} presents several inserting constructions and  multilevel constructions, based on the mixed dimension construction \cite{MixDD}.
We
first introduce the concept of SC-representation of MDCs to redescribe the mixed dimension construction \cite{MixDD}. Lemmas \ref{Trivial} and \ref{zhunze} introduce some auxiliary subspaces  
and present a sufficient condition for a $k$-dimensional subspace of $\BF_q^n$ to be added to the code from the mixed dimension construction. Theorem \ref{Insert} gives an inserting construction based on the mixed dimension construction. Furthermore, when the parameters in Theorem  \ref{Insert} satisfy some additional constraints, Theorems 
\ref{Insert2} and \ref{Insert2'} propose two inserting constructions based on Theorem \ref{Insert}. Subsequently, Theorem \ref{M-Insert} shows that the mixed dimension construction can still be combined with the multilevel construction.
 Our inserting mixed dimension constructions are further improved by combining with the multilevel construction in Theorems \ref{M-Insert2} and \ref{M-Insert2'}. In the multilevel constructions, rank-restricted  Ferrers diagram rank-metric codes play an important role. Applying Theorems \ref{M-Insert}, \ref{M-Insert2}, and \ref{M-Insert2'}, we establish new lower bounds for CDCs (see Theorems \ref{T-Mul-Mix}, \ref{T-Mul-Intert2}, and \ref{T-Multi-Insert2'}).

  Section \uppercase\expandafter{\romannumeral4} provides new lower bounds for CDCs and compares them with the previously best-known results. Corollaries \ref{C-Mulmixuseless}, \ref{2-12}, and \ref{q-12} present lower bounds on the size of $(2k,2\delta,\{k\})_q$, $(12+h,4,\{4\})_2$, and $(6\delta+h,2\delta,\{2\delta\})_q$ CDCs for general parameters. Additionally, Corollaries \ref{NEW2} and \ref{New3} provide explicit lower bounds for $(n,4,\{5\})_q$ CDCs with $15 \leq n \leq 19$ and $(n,4,\{6\})_q$ CDCs with $n=18, 19$, which depend on the selection of certain parameters. Overall at least 63 new lower bounds are given and compared with the results in \cite{MixDD}.


\section{Preliminaries}
In this section, we will review some fundamental definitions and known results that are essential for deriving our conclusions.
\subsection{Rank-Metric Codes}
We first recall the concept of rank-restricted rank-metric codes (RRMCs), which can be regarded as a generalization of rank-metric codes.

Let $m,n,$ and $d$ be positive integers, and let $r \in \{0,1, \cdots, \min \{m,n\}\}$. An $(m \times n, d)_{q}$ rank-metric code is called an $(m \times n, d; r)_{q}$ RRMC  if the rank of each codeword is at most $r$. The rank distribution of a linear MRD code can be determined from its parameters. The subsequent theorem is crucial for deriving lower bounds on RRMCs.

\begin{theorem}[Delsarte Theorem \cite{DELSARTE}]\label{T-rank-distribution}
Let $m, n, d$, and $i$ be positive integers with $d \leq i \leq \min\{m,n\}$. Then the number of codewords with rank $i$ in an $[m \times n, d]_q$ MRD code is given by
				\[D(m,n,d,i)_q := \begin{bmatrix}
													\min \{m, n\} \\
													i
									\end{bmatrix}_{q}
				\sum_{j = 0}^{i - d}(-1)^{j} q^{\frac{j(j-1)}{2}}
				 \begin{bmatrix}
													i \\
													j
									\end{bmatrix}_{q}
				(q^{\max{\{m,n\}}(i - d - j + 1)} - 1).
				\]				
\end{theorem}

Given an $[m \times n, d]_q$ MRD code $\cM$, we can construct an $(m \times n, d; r)_q$ RRMC using elements from $\cM$ that have ranks at most $r$. Consequently, a lower bound on the maximum size of an $(m \times n, d; r)_q$ RRMC can be expressed as $\Delta(m, n, d; r)_q := 1 + \sum_{i=d}^r D(m,n,d,i)_q$.

The following lemma provides a technique for constructing disjoint MRD subcodes from a linear MRD code, which is crucial for the inserting construction of CDCs.

\begin{lemma}[Subcode Construction \cite{Subcode} ]\label{L-Subcode}
Let $\cM$ be an $[m \times n, d]_q$ MRD code that contains an $[m \times n, d_1]_q$ MRD subcode $\cM_1$, where $d_1 > d.$ Let $s = \frac{\Delta(m, n, d)_q}{\Delta(m, n, d_1)_q}$. 
Then there exist $s$ subcodes of $\cM$ satisfying the following conditions:
\begin{itemize}
\item[(1)] $\cM_i$ is an $(m \times n, d_1)_q$ MRD code for all $1 \leq i \leq s$;
\item[(2)] for $1 \leq i < i' \leq s$, we have $\bM \neq \bM'$ and $\rank(\bM - \bM') \geq d$ for all $\bM \in \cM_i$, $\bM' \in \cM_{i'}$.
\end{itemize}
Moreover, $\cM_1$ is the unique linear MRD code among these $s$ subcodes.
\end{lemma}

\subsection{Ferrers Diagram Rank-Metric Codes}
Etzion and Silberstein introduced a new concept known as Ferrers diagram rank-metric code (FDRMC)  in \cite{Multilevel}.
 
 Given positive integers $m$ and $n$, an $m \times n$ Ferrers diagram $\cF$ is an $m \times n$ array consisting of dots and empty cells, where all the dots are positioned to the right within the diagram, the number of dots in each row does not exceed the number of dots in the previous row, and the first row contains $n$ dots and the rightmost column contains $m$ dots.

 We denote the number of dots in the $i$-th column of the Ferrers diagram $\cF$ by $\gamma_i$, where $1 \leq i \leq n$. Given positive integers $m$ and $n$, along with a sequence of integers satisfying $1 \leq \gamma_1 \leq \gamma_2 \leq \cdots \leq  \gamma_{n} = m$, there exists a unique Ferrers diagram $\cF$, where each $i$-th column has $\gamma_i$ dots for all $1 \leq i \leq n$. 
In this context, we denote $\cF$ by $[\gamma_1, \gamma_2, \dots,\gamma_{n}]$, and the transposed Ferrers diagram by
$\cF^\mathrm{T} =[\rho_{m}, \rho_{m-1}, \dots, \rho_1]$, where $\rho_i$ represents the number of dots in the $i$-th row of $\cF$, for $1 \leq i \leq m$. Note that a Ferrers diagram $\cF$ is called empty if $\cF$ contains no dots, and is called full if $\cF=[\gamma_1, \gamma_2, 
\dots, \gamma_{n}]$ with $\gamma_1=\gamma_2=\cdots=\gamma_{n}=m$.
\begin{example}
    Let $\cF=[1,1,2,3]$. Then 
    \[\cF=\left(\begin{array}{cccc} 
\bullet & \bullet &\bullet &\bullet\\
 & & \bullet &\bullet\\
 &&&\bullet
    \end{array}\right),~~  
 \cF^{\mathrm{T}}=\left(\begin{array}{ccc}
    \bullet&\bullet&\bullet\\
    &\bullet&\bullet\\
    &&\bullet\\
    &&\bullet 
    \end{array}\right).\] 
\end{example}

\begin{definition}\label{D-FRDMC}
Given an $m \times n$ Ferrers diagram $\cF$, an $(\cF, d)_q$ Ferrers diagram rank-metric code (FDRMC)  is an $(m \times n, d)_q$ rank-metric code  in which, for each matrix,  all entries that do not correspond to dots in the Ferrers diagram $\cF$ are zeros. In addition, an $(\cF, d)_q$ FDRMC with $M$ codewords is written as an $(\cF, M, d)_q$ FDRMC. 
\end{definition}
\begin{remark}
 If $\cF$ is a full Ferrers diagram, then its corresponding FDRMC is simply a classical rank-metric code. If there exists an $(\cF, M, d)_q $ FDRMC, then there also exists an $(\cF^\mathrm{T}, M, d)_q$ FDRMC.
\end{remark}
The following lemma provides a Singleton-like bound for FDRMCs.

 \begin{lemma}[\cite{Multilevel}]\label{L-Singleton}
Given an $(\cF, M, d)_q$ Ferrers diagram rank-metric code, let $v_i$ denote the number of dots in $\cF$ after removing the top $i$ rows and the rightmost $d - 1 - i$ columns, for $0 \leq i \leq d - 1$. Then the upper bound on $M$ is $q^{\min_i\{v_i\}}.$
\end{lemma}

An FDRMC that attains this bound is referred to as an optimal FDRMC. Constructions of optimal FDRMCs can be found in \cite{Opt-FDRMC,Multilevel,opt-F1,opt-F2,opt-F3}. The following theorem presents a universal method for verifying the existence of optimal FDRMCs.

\begin{theorem}[\cite{Opt-FDRMC}]\label{T-opt-FDRMC}
Assume $\cF$ is an $m \times n$ $(m \geq n)$ Ferrers diagram,  and each of the rightmost $d - 1$ columns has at least $n$ dots. Then there exists an optimal $(\cF, q^{\sum_{i = 1}^{n - d + 1}\gamma_{i}}, d)_q$ FDRMC, where $\gamma_i$ is the number of dots in the $i$-th column of $\cF$.
\end{theorem}
Obviously, optimal FDRMCs  always exist when $d \leq 2.$

Let $\sigma(\cdot)_{a,b}: \mathbb{F}_{q}^{m \times n} \rightarrow \mathbb{F}_{q}^{a \times b}$ be the map from a matrix $\bM$ to the submatrix of $\bM$ formed by the intersection of its top $a$ rows and rightmost $b$ columns. 

\begin{definition} Let $\cF$ be an $m \times n$ Ferrers diagram.
For non-negative integers $a$, $b$, and $r$ with $a \leq m$ and $b \leq n$,  
an $(\cF, d; a, b, r)_{q}$ rank-restricted Ferrers diagram rank-metric code (RFDRMC) $\cC_{\cF}$ is an $(\cF, d)_q$ FDRMC  such that the rank of $\sigma(\bA)_{a, b}$ is at most $r$ for each element $\bA \in \cC_{\cF}$. In addition, an $(\cF, d; a, b, r)_{q}$ RFDRMC with $M$ codewords is written as an $(\cF, M, d; a, b, r)_{q}$ RFDRMC.
\end{definition}

A straightforward construction for RFDRMCs was presented in \cite{MulLink}.

\begin{proposition}[\cite{MulLink}]\label{P-RFDRMC}
Let $m$, $n$, $d$, $a$, $b$ be positive integers with $m > a$ and $n > b$. Let $r \geq 0$.
Assume $\cF = \begin{pmatrix}			\cF_1 & \cF_2 \\
& \cF_3
\end{pmatrix}$ 
is an $m \times n$ Ferrers diagram, where $\cF_2$ is an $a \times b$ full Ferrers diagram, and $\cF_1$, $\cF_3$ are with compatible sizes.
Let $\cD_2$ be an $(a \times b, d; r)_q$ RRMC, and let $\cD_i$ be $(\cF_i, |\cD_i|, d)_q$ FDRMCs for $i = 1,3$.
Define
\[\cC_{\cF} =
	\Bigg\{\begin{pmatrix}
	\bD_1 & \bD_2\\	\bO		 & \bD_3
\end{pmatrix}: \bD_i \in \cD_i, i = 1, 2, 3
\Bigg\},
\]  then $\cC_{\cF}$ is an $(\cF, \prod_{i = 1}^{3}|\cD_i|,d;a, b, r)_q$ RFDRMC. 
\end{proposition}
\begin{remark}
For Ferrers diagrams $\cF$ with shape $(\cF_1~\cF_2)$ or $\begin{pmatrix}
    \cF_2\\
    \cF_3
\end{pmatrix}$, where $\cF_2$ is a full Ferrers diagram, the corresponding RFDRMCs can be derived using a similar method as in Proposition 
\ref{P-RFDRMC}.
\end{remark}

\subsection{Multilevel Construction}
Etzion and Silberstein \cite{Multilevel}  introduced the multilevel construction for constant dimension codes by establishing connections between subspace distance and Hamming distance or rank distance.

Let  $U$ be a subspace in $\mathcal{G}_{q}(n,k)$. It can be represented by a $k \times n$ generator matrix $\mathbf{U}$ whose rowspace forms $U$.  The distance between  two $k$-dimensional subspaces $U$ and $V$ of $\mathbb{F}_{q}^{n}$ is also given by 
\begin{align}\label{DS}
d_{S}(U, V) = 2 \cdot \text{rank} 
\begin{pmatrix}
\mathbf{U} \\
\mathbf{V}
\end{pmatrix} - 2k,
\end{align}
where $\mathbf{U}$ and $\mathbf{V}$ are generator matrices of subspaces $U$ and $V$, respectively. It is clear that the generator matrix of a subspace $U$ is not unique. However, there exists exactly one matrix in \emph{reduced row echelon form} (RREF) that can be obtained by applying the Gaussian elimination algorithm to any generator matrix of $U$. This matrix is denoted by $\xi(U)$. The identifying vector of $U$, denoted by $i(U)$, is the vector in $\BF_2^n$ where the ones  correspond to the pivots of $\xi(U)$. The Hamming weight of a vector is represented as $\wt(\cdot)$.

The Ferrers tableaux form of $U$, written as $\cF(U)$, is obtained from $\xi(U)$ through the following three steps:
\begin{itemize}
\item[(1)] for each row of $\xi(U)$, remove all zeros to the left of the pivot element;
\item[(2)] remove the pivot columns of $\xi(U)$;
\item[(3)] shift all remaining entries to the right.
\end{itemize}

The Ferrers diagram of $U$, denoted as $\cF_U$, is defined by replacing all entries in $\cF(U)$ with dots ``$\bullet$''.

The echelon Ferrers form of a vector $\mathbf{v}$ of length $n$ and  weight $k$, denoted by $EF(\mathbf{v})$, is the $k \times n$ matrix in RREF with leading entries (of rows) in the columns indexed by the nonzero entries of $\mathbf{v}$ and ``$\bullet$'' in  all entries that do not have terminals zeros or ones. Remove all zeros to the left of the pivot element in each row of $EF(\bv)$. Then remove the columns containing the pivot element. Finally, shift all remaining entries to the right. By following these steps, we obtain the Ferrers diagram of $EF(\mathbf{v})$, denoted as $\mathcal{F}_{\mathbf{v}}$.

 Actually, the Ferrers diagram of $U$ is identical to the Ferrers diagram of $EF(i(U))$.
\begin{example}\label{E-FDRMC}
Consider a subspace $U \in \mathcal{G}_{2}(8,3)$ with the generator matrix in RREF
\[\xi(U) = \begin{pmatrix}
						1 & 1 & 0 & 0 & 0 & 1 & 0 & 0\\
						0 & 0 & 1 & 0 & 0 & 1 & 1 & 0\\
						0 & 0 & 0 & 0 & 1 & 0 & 0 & 1 \\
					\end{pmatrix}.\]
The Ferrers tableaux form of $U$ is
\[
\mathcal{F}(U) = \begin{pmatrix}
						 1 &  0 &  1 & 0 & 0\\
						   &  0 &  1 & 1 & 0\\
						   &    &  0 & 0 & 1 \\
					\end{pmatrix}.
\]
The identifying vector $i(U)$ is $(10101000)$, and its echelon Ferrers form is
\begin{align*}
	EF(i(U)) = 
	\begin{pmatrix}
	1 & \bullet & 0 & \bullet & 0 & \bullet & \bullet & \bullet\\
	0 & 0 &       1 & \bullet & 0 & \bullet & \bullet & \bullet\\
	0 & 0 & 0 & 0 & 1 & \bullet & \bullet &\bullet \\
	\end{pmatrix}.
\end{align*}
The Ferrers diagram of $U$ and the Ferrers diagram of $EF(i(U))$ are identical, as shown below
 \begin{align*}
\cF_U=\cF_{i(U)} = 
	\begin{pmatrix}
	 \bullet  & \bullet  & \bullet & \bullet & \bullet\\
	       & \bullet & \bullet & \bullet & \bullet\\
	      &   & \bullet & \bullet &\bullet \\
	\end{pmatrix}.
\end{align*}
\end{example}

The following key result clarifies the relationship between the subspace distance of a pair of subspaces and the Hamming distance of their corresponding identifying vectors.

\begin{lemma}[\cite{Multilevel}]\label{L-S-H}
Let $X$ and $Y$ be two subspaces in $\mathcal{P}_{q}(n)$. Then  \[d_{S}(X,Y) \geq d_{H}(i(X), i(Y)),\] where $d_{H}(\cdot)$ denotes the Hamming metric.
\end{lemma}

Let $\cF$ be a Ferrers diagram and $\cC_{\cF}$ the
 corresponding $(\cF,d)_q$ 
 FDRMC. For a codeword $\bA \in \cC_{\cF}$, let $\bA_{\cF}$ denote the portion of $\bA$ that corresponds to the entries of $\cF$ in $\bA$.

\begin{definition}[\cite{opt-F3}]\label{D-LFRDMC}
Given a Ferrers diagram rank-metric code $\cC_{\cF} \subseteq \BF_q^{k \times (n- k)}$, a lifted Ferrers diagram rank-metric code $\mathscr{L}(\cC_{\cF})$ is defined as 
\[
\mathscr{L}(\cC_{\cF}) = \{X \in \cG_{q}(n,k): \cF(X) = \bA_{\cF}, \bA \in \cC_{\cF}\}.
\]
\end{definition}
The following theorem establishes the relationship between FDRMCs and CDCs.
\begin{theorem}[\cite{opt-F3}]\label{T-LFRDMC}
If $\cC_{\cF} \subseteq \BF_q^{k \times (n- k)}$ is an $(\cF, M, d)_q$ Ferrers diagram rank-metric code, then its lifted code $\mathscr{L}(\cC_{\cF})$ is an $(n, M, 2d,\{k\})_q$ constant dimension code.
\end{theorem}
The multilevel construction \cite{Multilevel}  for CDCs is based on Lemma \ref{L-S-H} and Theorem \ref{T-LFRDMC}.
 
\emph{Multilevel Construction}:
Initially, a binary constant-weight code of length $n$, weight $k$, and Hamming distance $2\delta$ is chosen as the set of identifying vectors for $\cC$. Subsequently, for each identifying vector, a corresponding lifted FDRMC with minimum subspace distance $2\delta$ is constructed. The union of these lifted FDRMCs forms an $(n, 2\delta, \{k\})_q$ CDC.

\subsection{Mixed Dimension/Distance Codes}
This subsection introduces the concept and constructions of a new class of subspace codes, known as mixed dimension/distance codes.

\begin{definition}\label{D-MDDC}
    Let $\cC \subseteq \cP_q(n)$. Suppose $d_1$ and $d_0$ are positive integers with $d_1 \geq d_0$. If for any two distinct subspaces $X,Y \in \cC$, $d_S(X,Y) \geq d_1$ when $\dim(X) = \dim(Y)$, and $d_S(X,Y) \geq d_0$  when $\dim(X) \neq \dim(Y)$, then we say that $\cC$ is an $(n,d_1,d_0)_q$ mixed dimension/distance code (MDDC).

     Let $T$ be a subset of $\{0, 1, \dots, n\}$. We refer to $\cC$ as an $(n, d_1, d_0, T)_q$ MDDC if $\eta_i(\cC) >0$ for $i \in T$ and $\eta_i(\cC) = 0$ for $i \notin T$. Additionally, an $(n,d_1,d_0,T)_q$ MDDC with $M$ codewords is denoted as an $(n, M, d_1, d_0,T)_q$ MDDC.
\end{definition}

\begin{example}\label{Alg}
    By using the $(8,4801,4,\{4\})_2$ CDC and the $(8,1326,4,\{3\})_2$ CDC constructed in \cite{Alg}, and by Algorithm 1 in \cite{MixDD}, we can obtain an $(8,5128,4,3,\{4,3\})_2$ MDDC $\cC$ with $\eta_4(\cC)=4801$ and $\eta_3(\cC)=327$.
\end{example}

The following proposition provides an explicit construction of an $(n,2\delta,\delta)_q$ MDDC. 

\begin{proposition}[\cite{MixDD}]\label{P-ConstrMDDC}
    Let $n$ and $\delta$ be two positive integers, and let $m=\lfloor \frac{n-1}{\delta}\rfloor$. For all integers $0 \leq i \leq m$, let $k_i=n-i\delta$,  and $\cC_i$ be an $(n,|\cC_i|,2\delta,\{k_i\})_q$ CDC. Then $\cC=\bigcup_{i=0}^m \cC_i$ is an $(n,2\delta,\delta,\{k_0,k_1,\dots,k_m\})_q $ MDDC with $\eta_{k_i}(\cC) = |\cC_i|$.
\end{proposition}

 The following theorem gives a lower bound on the size of an $(n, 2\delta, \delta + 1)_q$ MDDC. 
 
 \begin{theorem}[\cite{MixDD}] \label{T-ConstrMDDC}
     Given integers $n, k$, and $\delta$, where $\delta \geq 2$ and $n > k \geq 2\delta -1$. From an $(n, |\cC_0|, 2\delta, \{k\})_q$ CDC $\cC_0$, we can construct an $(n, 2\delta, \delta+1)_q$ MDDC $\cC$ with $\eta_k(\cC) = |\cC_0|$ and $\eta_{k-\delta+1}(\cC)= \max\{N,0\}$, where
     \[ N= \left\lfloor 
     \frac{[\begin{smallmatrix}
         n\\ k-\delta+1
     \end{smallmatrix}]_q - |\cC_0|[\begin{smallmatrix}
         k\\\delta-1
         \end{smallmatrix}]_q}{\sum_{i=0}^{\delta-1}q^{i^2}[\begin{smallmatrix}
             k-\delta+1\\i
         \end{smallmatrix}]_q [\begin{smallmatrix}
             n-(k-\delta+1)\\ i
         \end{smallmatrix}]_q} \right\rfloor.\]
 \end{theorem}

\section{Our constructions of constant dimension codes}

In this section, we begin by revisiting some basic concepts and the mixed dimension construction introduced in \cite{MixDD}. Subsequently, we present several inserting constructions and multilevel constructions based on the mixed dimension construction.

\subsection{Mixed Dimension Construction}
To present our results, we will use the following notations.
\begin{itemize}
    \item $\rs(\bU)$ denotes the subspace of $\BF_q^n$ spanned by the rows of  a matrix $\bU \in \BF_q^{k \times n}$,
\item $\bI_k$ denotes the identity matrix in $\BF_q^{k \times k}$,
\item $\bO_{a \times b}$ denotes the all zero matrix of size $a\times b$, and we will omit the size when it is evident from the context,
\item $(\bA~|~\bB)$ or $(\bA~\bB)$  represents a matrix concatenated from $\bA$ and $\bB$ with compatible sizes and ambient fields, 
\item we denote \[N_q(x,y,z)= \left\lfloor 
    \frac{[\begin{smallmatrix}
         x\\ z-y+1
     \end{smallmatrix}]_q - S_q(x,2y,z)[\begin{smallmatrix}
         z\\ y-1
         \end{smallmatrix}]_q}{\sum_{i=0}^{y-1}q^{i^2}[\begin{smallmatrix}
             z-y+1\\ i
         \end{smallmatrix}]_q [\begin{smallmatrix}
             x-(z-y+1)\\ i
         \end{smallmatrix}]_q} \right\rfloor,\]
where $S_q(x,2y,z)$ represents the best-known size of the $(x,2y, \{z\})_q$ CDC,
\item for a subset $T$ of the positive integers, denote the minimum value in the set $T$  by $    T^{\min}$, and define
\begin{align*}
    l_T:= \begin{cases}
        \min\{u-v:u,v \in T, u>v\} & \text{if}~|T|>1, \\
        $0$ & \text{if}~ |T|=1.
    \end{cases}
\end{align*}

\end{itemize}

To re-present the mixed dimension construction from \cite{MixDD}, we extend the concept of the SC-representation of CDCs to MDCs.
\begin{definition} \label{D-SC}
A set of matrices $\cH(\cC) \subseteq \BF_q^{k \times n}$ is called  an SC-representation of a constant dimension code $\cC$ in $\cG_q(n,k)$ if the following conditions are satisfied.
\begin{itemize} 
\item[(1)] For each $\bH \in \cH(\cC)$, $\rank(\bH) = k$.
\item[(2)] For any two distinct elements $\bH, \bH' \in \cH(\cC)$, $\rs(\bH) \neq  \rs(\bH')$.
\item[(3)] $\cC = \{\rs(\bH) \mid \bH \in \cH(\cC)\}$.
\end{itemize}
Furthermore, let $\cC$ be an $(n, 2\delta, T)_q$ mixed dimension code such that $ \cC= \bigcup_{k\in T}\cC^{k}$ with $\cC^{k} \subseteq \cG_q(n,k)$. 
An SC-representation of the mixed dimension code $\cC$ is defined as $\cH(\cC) = \bigcup_{k \in T} \cH(\cC^{k})$.
\end{definition}

The subsequent theorem provides the mixed dimension construction of CDCs from MDDCs and RMCs.

\begin{theorem}[Mixed dimension construction \cite{MixDD}]\label{Mix}
    Let $n, n_1, n_2, k,$ and $ \delta$ be integers with $n=n_1+n_2$, $n_1 \geq k$, $n_2 \geq k$, and $k \geq \delta \geq 2$.  Let $T_1 \subseteq [\delta,k]$ with $l_{T_1} <2 \delta$. Let $T_2 \subseteq [k+\delta-T_1^{\min},k]$ with $l_{T_2} < 2\delta$.   For $i = 1,2$, let $\cH_i = \bigcup_{r \in T_i}\{\bH \in \BF_q^{r \times n_i}: \rank(\bH)=r, \bH~\text{in}~\RREF \}$ be an SC-representation of  an $(n_i, |\cX_i|, 2\delta, 2\delta-l_{T_i}, T_i)_q$ MDDC $\cX_i$. Let $\cP_t$ be a $(k \times (n_2+t-k),\delta)_q$ MRD code for all $t \in T_1$ and $\cQ_s$ be a $(k \times (n_1+s-k), \delta; T_1^{\min} - \delta -(k-s))_q$ RRMC for all $s\in T_2$.
    Define $\cC_1 = \bigcup_{t \in T_1} \cC_1^{(t)}$ and $\cC_2 = \bigcup_{s \in T_2} \cC_2^{(s)}$,
    where
    \begin{align*}
        \cC_1^{(t)} &= \left\{
        \rs\left(
        \begin{array}{c|cc}
        \begin{matrix}
            \bH_1 \\ \bO   
        \end{matrix}
        &\begin{matrix}
            \bO \\
            \bI_{k-t}
        \end{matrix}
            & \bP
        \end{array}
        \right): \bH_1 \in \cH_1, \rank(\bH_1) = t, \bP \in \cP_t
        \right\},\\   \cC_2^{(s)} &=  \left\{\rs\left( 
        \begin{array}{cc|c}  
            \bQ &
      \begin{matrix}
                \bO \\
          \bI_{k-s} 
        \end{matrix}
        &
    \begin{matrix}
            \bH_2\\
            \bO
        \end{matrix}
        \end{array}
        \right)
        :\bQ \in \cQ_s, \bH_2 \in \cH_2, \rank(\bH_2)=s
        \right\}.
    \end{align*}
    Then  $\cC_1 \cup \cC_2$ is an $(n, 2\delta, \{k\})_q$ CDC with the size of  $\sum_{t \in T_1} \eta_t(\cX_1)|\cP_t| + \sum_{s \in T_2} \eta_s(\cX_2)|\cQ_s|$.
\end{theorem}

\subsection{Inserting Construction}

 Additional $k$-dimensional subspaces can be inserted into the codes from the mixed dimension construction if they possess unique properties.
 
 We first demonstrate that there exist special subspaces $K_1^{(t)}$ and $K_2^{(s)}$ of $\BF_q^n$ for all $t \in T_1$ and $s \in T_2$. 
 These subspaces intersect trivially with the subspaces in $\cC_1^{(t)}$ and $\cC_2^{(s)}$, respectively, only at the zero vector. This is stated in the following lemma.
\begin{lemma}\label{Trivial}
    For  $\cC_1^{(t)}$ and $ \cC_2^{(s)}$ constructed  as in Theorem \ref{Mix}, there exists an $(n_2-k+t)$-dimensional subspace $K_1^{(t)}$ of $\BF_q^n$ that intersects trivially with the codewords of $\cC_1^{(t)}$, and there exists an $(n_1-k+s)$-dimensional subspace $K_2^{(s)}$ of $\BF_q^n$ that intersects trivially with the codewords of $\cC_2^{(s)}$, for all $t \in T_1$, $s \in T_2.$
\end{lemma}
\begin{IEEEproof}
    Set \[K_1^{(t)} = \rs(\bO_{(n_2-k+t) \times (n_1+k-t)}~|~\bI_{n_2-k+t})\] and \[K_2^{(s)} = \rs(\bI_{n_1-k+s}~|~\bO_{(n_1-k+s) \times (n_2+k-s)}).\]
    For $U_1^{(t)} \in \cC_1^{(t)}$, we have
    \begin{align*}   \dim(U_1^{(t)}+K_1^{(t)})&=\rm{rank}\left(  \begin{array}{c:c:c}
        \begin{matrix}
            \bH_1 \\
            \bO  
        \end{matrix}& \begin{matrix}
            \bO \\
            \bI_{k-t} 
        \end{matrix} &
        \bP\\
        \hdashline
            \bO_{(n_2-k+t) \times n_1} & \bO_{(n_2-k+t) \times (k-t)} & \bI_{n_2-k+t}  
        \end{array}\right)\\
        &=\rank(\bH_1)+(k-t)+(n_2-k+t)=t+n_2,
    \end{align*}
    where $\bH_1 \in \cH_1, \rank(\bH_1) = t,$ and $\bP \in \cP_t$.
    For $U_2^{(s)} \in \cC_2^{(s)}$, we have
    \begin{align*}
        \dim(U_2^{(s)}+K_2^{(s)})&= \rm{rank}
        \left(  \begin{array}{c:c:c}
                \bQ & \begin{matrix}
                    \bO \\
                    \bI_{k-s} 
              \end{matrix}&
              \begin{matrix}
                  \bH_2\\
                  \bO
              \end{matrix}\\
                \hdashline
            \bI_{n_1-k+s}&\bO_{(n_1-k+s) \times (k-s)} & \bO_{(n_1-k+s) \times n_2}
        \end{array}\right)\\
        &= \rank(\bH_2)+(k-s)+(n_1-k+s)
        = s+n_1,
    \end{align*}
    where $\bH_2 \in \cH_2$, $\rank(\bH_2)=s,$ and $\bQ \in \cQ_s$. Therefore, for all $t \in T_1$ and $s \in T_2$,  we have
    \begin{align*}
        \dim(U_1^{(t)} \cap K_1^{(t)}) &= \dim(U_1^{(t)}) + \dim(K_1^{(t)}) - \dim(U_1^{(t)}+K_1^{(t)})\\
        &= k +(n_2-k+t)-(t+n_2) = 0,\\
        \dim(U_2^{(s)} \cap K_2^{(s)}) &= \dim(U_2^{(s)}) + \dim(K_2^{(s)}) - \dim(U_2^{(s)}+K_2^{(s)})\\ &=k+(n_1-k+s)-(s+n_1)=0.
    \end{align*}
\end{IEEEproof}

Based on Lemma \ref{Trivial}, we provide a sufficient condition for adding a codeword in $\cG_q(n,k)$ to the code obtained from the mixed dimension construction.

\begin{lemma}\label{zhunze}
    With the same notations used in Lemma \ref{Trivial},  if $B$ is a $k$-dimensional subspace of $\BF_q^n$ satisfying $\dim(B \cap K_1^{(t)}) \geq \delta$ and $\dim(B \cap K_2^{(s)}) \geq \delta$, then $d_S(B,U_1^{(t)}) \geq 2\delta$ and $d_S(B,U_2^{(s)}) \geq 2\delta$ for all codewords $U_1^{(t)} \in \cC_1^{(t)}$ and $ U_2^{(s)} \in \cC_2^{(s)}$, where $t \in T_1$ and $s \in T_2$.
\end{lemma}
\begin{IEEEproof}
        For all $U_1^{(t)} \in \cC_1^{(t)}$ and $t \in T_1$, it is clear that
    \begin{align*}
    \dim(B \cap U_1^{(t)}) &\leq \dim((B \cap K_1^{(t)}) \cap U_1^{(t)}) + k -\delta \\
    &= \dim(B \cap (K_1^{(t)} \cap U_1^{(t)}))+k-\delta \\ &= k-\delta.
    \end{align*}
    Similarly, we have $\dim(B \cap U_2^{(s)}) \leq k - \delta$ for all $U_2^{(s)} \in \cC_2^{(s)}$ and  $s \in T_2$. 
    
   It follows that $d_S(B,U_1^{(t)}) \geq 2\delta$ and $d_S(B,U_2^{(s)}) \geq 2\delta$.
\end{IEEEproof}

Let $k$ and $m$ be positive integers with $k \leq m$. Let $\mathbf{A} \in \mathbb{F}_{q}^{k \times m}$ be a full-rank matrix in RREF. 
Define the embedding map $\tau_{\bA}: \BF_q^{k' \times (m - k)} \rightarrow \mathbb{F}_{q}^{k' \times m}$ by inserting $k$ zero columns $(0,0, \dots, 0)^{\mathrm{T}}$ of length $k'$ into $\bB$ in the positions corresponding to the pivot columns of $\mathbf{A}$, where $\bB \in \BF_q^{k' \times (m - k)}$.
For example,
	\begin{align*}
	\mathbf{A} = \begin{pmatrix}
					1 & 1 & 0 & 0 & 0 & 0 & 1 \\
					0 & 0 & 1 & 1 & 0 & 0 & 1 \\
					0 & 0 & 0 & 0 & 0 & 1 & 1 \\
	    \end{pmatrix} \in \mathbb{F}_{2}^{3 \times 7} \ \text{and} \
	\mathbf{B} = \begin{pmatrix}
	0 & 1 & 0 & 1\\
	1 & 1 & 0 & 0\\
	0 & 0 & 1 & 0\\
	0 & 0 & 1 & 1\\ 
	\end{pmatrix} \in \mathbb{F}_{2}^{4 \times (7-3)},
	\end{align*}
where $\mathbf{A}$ has column positions $1,3,6$. Then
		\begin{align*}
		\tau_{\mathbf{A}}(\mathbf{B}) = 
		 \begin{pmatrix}
	0 & 0 & 0 & 1 & 0 & 0 & 1\\
	0 & 1 & 0 & 1 & 0 & 0 & 0\\
	0 & 0 & 0 & 0 & 1 & 0 & 0\\
	0 & 0 & 0 & 0 & 1 & 0 & 1\\ 
	\end{pmatrix} \in \mathbb{F}_{2}^{4 \times 7}.
		\end{align*}
  
Utilizing the embedding map $\tau_\bA$ described above, we propose an inserting construction based on the mixed dimension construction.
\begin{theorem}\label{Insert}
    Take the same notations used in Theorem \ref{Mix}.  Let $t_1, t_2, a_1,a_2,b_1,$ and $b_2$ be integers satisfying $ a_1+a_2=k,$ $ b_1+b_2 \geq \delta$,  $a_1 \leq t_1 \leq n_1-k+T_2^{\min}-\delta$, 
    $a_2+k-T_1^{\min} \leq t_2 \leq n_2- \delta$, and $a_i \geq \delta$,  $1 \leq b_i \leq \delta$ for $i = 1, 2$.  Let $\cA_1=\{\bA_1 \in \BF_q^{a_1 \times t_1}: \rank(\bA_1) = a_1, \bA_1~\text{in}~\RREF\}$ and $\cA_2=\{\bA_2 \in \BF_q^{a_2 \times (t_2-k+T_1^{\min})}: \rank(\bA_2) = a_2, \bA_2~\text{in}~\RREF\}$
    be SC-representations of two $(t_1, 2\delta, \{a_1\})_q$ and $(t_2-k+T_1^{\min}, 2\delta, \{a_2\})_q$ CDCs, respectively. 
    Let
    \begin{align*}
        \cM_2 &= \{(\bF_2~|~\bM_2): \bF_2 \in \BF_q^{a_2 \times (t_1-a_1)}, \bM_2 \in \BF_q^{a_2\times(n_1-k+T_2^{\min}-t_1)}\},\\   
        \cM_3 &= \{(\bF_3~|~\bM_3): \bF_3 \in \BF_q^{a_1 \times (t_2-k+T_1^{\min}-a_2)}, \bM_3 \in \BF_q^{a_1 \times(n_2-t_2)}\}
    \end{align*}
     be two RRMCs with parameters $(a_2 \times (n_1-k+T_2^{\min}-a_1), \delta; a_2-\delta)_q$ and $(a_1 \times (n_2-k+T_1^{\min}-a_2), \delta;a_1-\delta)_q$, respectively. For another integer $f$, let $\cM_1^r$ be an $(a_1 \times (n_1-k+T_2^{\min}-t_1), \delta)_q$ MRD code and $\cM_4^r$ be an $(a_2 \times (n_2-t_2), \delta)_q$ MRD code for all $1 \leq r \leq f$. Assume that $\bM \neq \bM'$ and $\rank(\bM -\bM') \geq b_1$ for $\bM \in \cM_1^r$ and $ \bM' \in \cM_1^{r'}$ for all $1 \leq r < r' \leq f$. Similarly, assume that $\bM \neq \bM'$ and $\rank(\bM -\bM') \geq b_2$ for $\bM \in \cM_4^r$ and $ \bM' \in \cM_4^{r'}$ for all $1 \leq r < r' \leq f$. Define $\cC_3 = \bigcup_{r=1}^f \cC_3^{(r)}$, 
    \begin{align*}
        \cC_3^{(r)}=\left\{
        \rm{rs}
        \begin{pmatrix}
            \bA_1 & \bM_1 & \bO_{a_1 \times(2k-T_1^{\min}-T_2^{\min})}& \tau_{\bA_2}(\bF_3) & \bM_3\\
            \tau_{\bA_1}(\bF_2) & \bM_2 & \bO_{a_2 \times(2k-T_1^{\min}-T_2^{\min})}&
            \bA_2 & \bM_4
        \end{pmatrix}
         \right\}, 
    \end{align*}
    where $\bA_i \in \cA_i$ for $i = 1,2$, $(\bF_i|\bM_i) \in \cM_i$ for $i = 2,3$, and $\bM_i \in \cM_i^r$ for $i = 1,4$.
     Then $\cC_1 \cup \cC_2 \cup \cC_3$ is an $(n, \sum_{i=1}^3 |\cC_i|, 2\delta, \{k\})_q$ CDC.
\end{theorem}
\begin{IEEEproof}
    First, we prove that $\cC_3$ is an $(n, 2\delta, \{k\})_q$ CDC.  It is straightforward to verify  that
      the elements of $\cC_3$ are $k$-dimensional subspaces of $\BF_q^n$. 
    Let
    \begin{align*}
        U_3= \rs(\bG_1) =
        \rs\begin{pmatrix}
            \bA_1 & \bM_1 & \bO_{a_1 \times(2k-T_1^{\min}-T_2^{\min})}&\tau_{\bA_2}(\bF_3) & \bM_3\\
            \tau_{\bA_1}(\bF_2) & \bM_2 &
            \bO_{a_2 \times(2k-T_1^{\min}-T_2^{\min})}&
            \bA_2 & \bM_4
        \end{pmatrix} \in \cC_3^{(r)}
    \end{align*}
     and
        \begin{align*}
        U_3'= \rs(\bG_2) =
        \rs\begin{pmatrix}
            \bA'_1 & \bM'_1 & \bO_{a_1 \times(2k-T_1^{\min}-T_2^{\min})}&\tau_{\bA'_2}(\bF'_3) & \bM'_3\\
            \tau_{\bA'_1}{(\bF'_2)} & \bM'_2 &\bO_{a_2 \times(2k-T_1^{\min}-T_2^{\min})}&
            \bA'_2 & \bM'_4
        \end{pmatrix} \in \cC_3^{(r')}
    \end{align*}
      be two distinct $k$-dimensional subspaces.
    The intersection of $U_3$ and $U_3'$ is
    \begin{align*}
        U_3 \cap U_3' = \{(\alpha_1, \alpha_2)\bG_1 = (\beta_1, \beta_2)\bG_2: \alpha_i, \beta_i \in \BF_q^{a_i}, i = 1,2\}.
    \end{align*}
    Now, we determine the minimum distance of $\cC_3$ in the following cases.
    \begin{itemize}
        \item[(1)] If $\bA_1 \neq \bA'_1$, then $\dim(U_3 \cap U_3') \leq \dim(\rs(\bA_1) \cap\rs(\bA'_1)) + a_2 \leq (a_1 - \delta) +a_2 $\\$\leq k -  \delta$. 
        \item[(2)] If $\bA_1 = \bA'_1$ and $\bA_2 \neq \bA'_2$, then $\dim(U_3 \cap U_3') \leq \dim(\rs(\bA_2) \cap \rs(\bA'_2))+ a_1 \leq $\\$ (a_2 - \delta) + a_1 \leq k - \delta$.
        \item[(3)] If $\bA_1 = \bA'_1$ and $\bA_2 = \bA'_2$,  then we obtain that
        \begin{align*}
        \begin{cases}
            \alpha_1 \bA_1+\alpha_2 \tau_{\bA_1}(\bF_2) =\beta_1 \bA_1+\beta_2 \tau_{\bA_1}(\bF'_2)
            \\
            \alpha_1\tau_{\bA_2}(\bF_3)+\alpha_2\bA_2=\beta_1\tau_{\bA_2}(\bF'_3)+\beta_2\bA_2
        \end{cases}.
        \end{align*}
    According to the definitions of $\tau_{\bA_1}(\bF_2)$ ,$\tau_{\bA_2}(\bF_3)$, $\tau_{\bA_1}(\bF'_2)$, and $\tau_{\bA_2}(\bF'_3)$, we can conclude that $\alpha_1 = \beta_1$ and $\alpha_2 = \beta_2$. Hence, we have 
    \begin{align*}
    \dim(U_3 \cap U_3') = \dim(\{(\alpha_1, \alpha_2): (\alpha_1, \alpha_2)\bG=\mathbf{0}, \alpha_i \in \BF_q^{a_i}, i =1,2\}),
    \end{align*}
    where $\bG=\begin{pmatrix}
        \bG_{11} & \bG_{12}\\
        \bG_{21} & \bG_{22}
    \end{pmatrix}$ with $\bG_{11}=(\bO|\bM_1-\bM'_1), \bG_{12} = (\tau_{\bA_2}(\bF_3)|\bM_3) -(\tau_{\bA_2}(\bF'_3)|\bM'_3),\bG_{21}=(\tau_{\bA_1}(\bF_2)|\bM_2)-(\tau_{\bA_1}(\bF'_2)|\bM'_2),$ and $\bG_{22} = (\bO|\bM_4-\bM_4')$.
    
    We continue to  analyze the following cases. 
    \begin{itemize}
        \item If $\bG_{12} \neq \bO$ or $\bG_{21} \neq \bO$, then 
        \begin{align*}
        \dim(U_3 \cap U_3') &\leq k - \max\{\rank((\bF_3|\bM_3)-(\bF_3'|\bM_3')), \rank((\bF_2|\bM_2)-(\bF'_2|\bM'_2))\} \\&\leq k-\delta.
            \end{align*}
            \item When $\bG_{12} = \bO$ and $\bG_{21} = \bO$, then
            \begin{align*}
                \dim(U_3 \cap U_3') &= \dim (\{\alpha_1\in \BF_q^{a_1}: \alpha_1\bG_{11}=\mathbf{0}\})+
                \dim (\{ \alpha_2 \in \BF_q^{a_2}: \alpha_2\bG_{22}=\mathbf{0}\})\\
                &= a_1 - \rank(\bO|\bM_1-\bM'_1) + a_2-\rank(\bO|\bM_4-\bM'_4)\\
                &=k-\rank(\bM_1-\bM'_1)-\rank(\bM_4-\bM'_4).
            \end{align*}
            \begin{itemize}
        \item[(\romannumeral1)]
        If $r = r'$, then $\bM_1, \bM'_1 \in \cM_1^r$ and $\bM_4, \bM'_4 \in \cM_4^r$. If $\bM_1 \neq \bM'_1$, then $\dim(U_3 \cap U_3') \leq k- \rank(\bM_1 - \bM'_1)\leq k - \delta$. If $\bM_1 = \bM'_1$, then $\bM_4 \neq \bM'_4$, and $\dim(U_3 \cap U_3') = k- \rank(\bM_4 - \bM'_4)\leq k - \delta$.
        \item[(\romannumeral2)] If $r \neq r'$, then
        \begin{align*}
          \dim(U_3 \cap U_3') &= k-\rank(\bM_1-\bM'_1)-\rank(\bM_4-\bM'_4) \\&\leq k-b_1-b_2 \leq k-\delta.
        \end{align*}
            \end{itemize}
    \end{itemize}
    \end{itemize}
    
    In conclusion, we have $\dim(U_3 \cap U_3') \leq k - \delta$. Therefore, $d_S(U_3, U_3') = 2k - 2\dim(U_3 \cap U_3') \geq 2\delta$.

    We need to prove that $\cC_1 \cup \cC_2 \cup \cC_3$ is an $(n, 2\delta, \{k\})_q$ CDC. Using the same notations as in Lemma \ref{Trivial}, it suffices to prove that $\dim(U_3 \cap K_1^{(t)}) \geq \delta$ and $\dim(U_3 \cap K_2^{(s)}) \geq \delta$  for all $t\in T_1$, $s \in T_2$, and for any codeword $U_3 \in \cC_3$. It is straightforward to verify that
    \begin{align*}
        \dim(U_3+K_1^{(t)})&=\rank \left(\begin{array}{c:c}
        \begin{matrix}
        \bA_1 & \bM_1 & \bO_{a_1\times(k-T_2^{\min})}\\ 
        \tau_{\bA_1}(\bF_2) &\bM_2 &\bO_{a_2\times(k-T_2^{\min})} 
        \end{matrix} &
        \begin{matrix}
           \bO_{a_1\times(k-T_1^{\min})}& \tau_{\bA_2}(\bF_3) & \bM_3\\
            \bO_{a_2\times(k-T_1^{\min})}&\bA_2 &\bM_4
        \end{matrix} \\
        \hdashline
        \bO_{(n_2-k+t) \times n_1} & \bO_{(n_2-k+t) \times (k-t)}~|~\bI_{n_2-k+t}
        \end{array}\right)\\
        &= a_1 + (n_2-k+t) + \rank(\bF_2|\bM_2),   
    \end{align*}
 where $\rank(\bF_2|\bM_2) \leq a_2 - \delta$. Similarly, we obtain that
    \begin{align*}
        \dim(U_3+K_2^{(s)})&=\rank
        \left(\begin{array}{c:c}
        \begin{matrix}
        \bA_1 & \bM_1 & \bO_{a_1\times(k-T_2^{\min})}\\
        \tau_{\bA_1}(\bF_2) &\bM_2 &\bO_{a_2\times(k-T_2^{\min})}
        \end{matrix} &
        \begin{matrix}
            \bO_{a_1\times(k-T_1^{\min})}&\tau_{\bA_2}(\bF_3) & \bM_3\\
            \bO_{a_1\times(k-T_1^{\min})}&\bA_2 &\bM_4
        \end{matrix} \\
        \hdashline
        \bI_{n_1-k+s}~|~\bO_{(n_1-k+s) \times (k-s)} & \bO_{(n_1-k+s) \times n_2}
     \end{array}\right)\\
        &= a_2 + (n_1-k+s) + \rank(\bF_3|\bM_3),
    \end{align*}
    where $\rank(\bF_3|\bM_3) \leq a_1-\delta$.
    Then, it is obvious that
    \begin{align*}
        \dim(U_3 \cap K_1^{(t)}) &= \dim(U_3) + \dim(K_1^{(t)}) - \dim(U_3+K_1^{(t)}) \\
        &\geq k+(n_2-k+t)-[a_1+(n_2-k+t) + a_2 - \delta] = \delta,\\
        \dim(U_3 \cap K_2^{(s)}) &= \dim(U_3) + \dim(K_2^{(s)}) - \dim(U_3+K_2^{(s)}) \\&\geq k+(n_1-k+s)-[a_2+(n_1-k+s)+a_1-\delta] = \delta.
    \end{align*}
    
    It follows that $\cC_1 \cup \cC_2 \cup \cC_3$ is an $(n,\sum_{i=1}^3 |\cC_i|, 2\delta,\{k\})_q$ CDC.  
\end{IEEEproof}

Furthermore, we consider the case where $n_2-t_2\geq a_2$, and provide two different constructions by inserting more subspaces into the CDCs presented in Theorem \ref{Insert}.

\begin{theorem}\label{Insert2}
    Take the same notations used in Theorem \ref{Insert}. Assume that $n_2-t_2 \geq a_2$ and $k-t_1\geq \delta$. Let $\cB_1 = \{\bB_1 \in \BF_q^{a_1 \times t_1}: \rank(\bB_1)=a_1, \bB_1 ~\text{in}~\RREF\}$ and $\cB_2=\{\bB_2 \in \BF_q^{a_2 \times (n_2-t_2)}: \rank(\bB_2)=a_2, \bB_2 ~\text{in}~\RREF\}$
    be SC-representations of two $(t_1, 2\delta, \{a_1\})_q$ and  $(n_2-t_2, 2\delta, \{a_2\})_q$ CDCs, respectively. Let $\cN_1$ be an $(a_1 \times (n_1 - k + T_2^{\min}-t_1), 
 \delta)_q$ MRD code and $\cN_4$ be an $(a_2 \times (t_2-k+T_1^{\min}), \delta; k - t_1 - \delta)_q$ RRMC. Let 
 \begin{align*}
     \cN_2=\left\{ (\bE_2~| ~\bN_2): \bE_2 \in \BF_q^{a_2 \times (t_1-a_1)},\bN_2 \in \BF_q^{a_2 \times (n_1 - k + T_2^{\min}-t_1)} 
\right\} 
\end{align*}
and
 \begin{align*}
     \cN_3=\left\{(\bN_3~|~\bE_3): \bN_3 \in \BF_q^{a_1 \times (t_2-k+T_1^{\min})}, \bE_3 \in \BF_q^{a_1 \times (n_2-t_2-a_2)} \right\}
 \end{align*}
be two $(a_2 \times (n_1 -k + T_2^{\min} -a_1), \delta; a_2-\delta)_q$ and $(a_1 \times (n_2-k+T_1^{min}-a_2), \delta; a_1-\delta)_q$ RRMCs, respectively. Define
\begin{align*}
    \cC_4 = 
    \left\{
    \rs\begin{pmatrix}
        \bB_1 & \bN_1 & \bO_{a_1 \times (2k-T_1^{\min}-T_2^{\min})} & \bN_3 & \tau_{\bB_2}(\bE_3)\\
        \tau_{\bB_1}(\bE_2) & \bN_2 &\bO_{a_2 \times (2k-T_1^{\min}-T_2^{\min})} & \bN_4 & \bB_2
    \end{pmatrix}
    \right\},
\end{align*}
where $\bB_i \in \cB_i$ for $i = 1,2$, $(\bE_2|\bN_2) \in \cN_2$, $(\bN_3 |\bE_3) \in \cN_3$, and $\bN_i \in \cN_i$ for $i=1,4$.
Then $\bigcup_{i=1}^4 \cC_i$ is an $(n, \sum_{i=1}^4 |\cC_i|, 2\delta, \{k\})_q$ CDC.
\end{theorem}
\begin{IEEEproof}
First, we prove that $\cC_4$ is an $(n, 2\delta, \{k\})_q$ CDC. It is straightforward to verify that any codeword in $\cC_4$ is a $k$-dimensional subspace of $\BF_q^n$. Let
\begin{align*}
    U_4 &= \rs(\bG_1) =
    \rs
    \begin{pmatrix}
        \bB_1 & \bN_1 & \bO_{a_1 \times (2k-T_1^{\min}-T_2^{\min})} & \bN_3 & \tau_{\bB_2}(\bE_3)\\
        \tau_{\bB_1}(\bE_2) & \bN_2 &\bO_{a_2 \times (2k-T_1^{\min}-T_2^{\min})} & \bN_4 & \bB_2
    \end{pmatrix}
\end{align*}
and 
\begin{align*}
    U_4' &= \rs(\bG_2) =
    \rs
    \begin{pmatrix}
        \bB'_1 & \bN'_1 & \bO_{a_1 \times (2k-T_1^{\min}-T_2^{\min})} & \bN'_3 & \tau_{\bB'_2}(\bE'_3)\\
        \tau_{\bB'_1}(\bE'_2) & \bN'_2 &\bO_{a_2 \times (2k-T_1^{\min}-T_2^{\min})} & \bN'_4 & \bB'_2
    \end{pmatrix}
\end{align*}
be two distinct $k$-dimensional subspaces in $\cC_4$. The intersection of $U_4$ and $U_4'$ is 
\begin{align*}
    U_4 \cap U_4' = \left\{
    (\alpha_1, \alpha_2)\bG_1=(\beta_1,\beta_2)\bG_2: \alpha_i,\beta_i \in \BF_q^{a_i},i=1,2
    \right\}.
\end{align*}

Now, we determine the minimum distance of $\cC_4$ from the following cases.
\begin{itemize}
    \item [(1)] If $\bB_1 \neq \bB'_1$, then $\dim(U_4 \cap U_4') \leq \dim(\rs(\bB_1) \cap \rs(\bB'_1)) + a_2 \leq (a_1-\delta)+a_2=k-\delta$.
    \item[(2)] If $\bB_1 = \bB'_1$ and $\bB_2 \neq \bB'_2$, then $\dim(U_4 \cap U_4') \leq \dim(\rs(\bB_2) \cap \rs(\bB'_2))+a_1 \leq (a_2 - \delta) +a_1=k-\delta$. 
    \item[(3)]If $\bB_1 = \bB'_1$ and $\bB_2 = \bB'_2$, we obtain that
    \begin{align*}
        \begin{cases}
            \alpha_1 \bB_1+\alpha_2 \tau_{\bB_1}(\bE_2) =\beta_1 \bB_1+\beta_2 \tau_{\bB_1}(\bE'_2)
            \\
            \alpha_1\tau_{\bB_2}(\bE_3)+\alpha_2\bB_2=\beta_1\tau_{\bB_2}(\bE'_3)+\beta_2\bB_2
        \end{cases}.
        \end{align*}
        
        According to the definitions of $\tau_{\bB_1}(\bE_2), \tau_{\bB_1}(\bE'_2)$, $\tau_{\bB_2}(\bE_3), $ and $\tau_{\bB_2}(\bE'_3)$, we can conclude 
 that $\alpha_1 = \beta_1$ and $\alpha_2 = \beta_2$. Hence,
        \begin{align*}
            \dim(U_4 \cap U_4')=\dim(\{(\alpha_1, \alpha_2): (\alpha_1, \alpha_2)\bG 
            =\mathbf{0}, \alpha_i \in \BF_q^{a_i}, i=1,2
            \}),
        \end{align*}
        where \[\bG=\begin{pmatrix}
                \bO_{a_1 \times (t_1-a_1)}  & \bN_1-\bN'_1 & \bN_3-\bN_3' & \bE_3-\bE'_3 \\
                \bE_2-\bE'_2 & \bN_2 - \bN'_2 & 
                \bN_4 -\bN'_4 &
                \bO_{a_2 \times (n_2 - t_2-a_2)}
        \end{pmatrix}.\]
        
Then
\begin{equation*}
\begin{aligned}
    \dim(U_4 \cap U_4') &\leq k - \rm{max} \{\rank(\bN_1-\bN'_1),\rank(\bN_4-\bN'_4),\\&~~~~\rank((\bE_2|\bN_2) -(\bE'_2|\bN'_2)),\rank((\bN_3|\bE_3) -(\bN'_3|\bE'_3)) \} \\&\leq k - \delta.
\end{aligned}
\end{equation*} 
\end{itemize}

In conclusion, we have $\dim(U_4 \cap U_4') \leq k - \delta$. It follows that $d_S(U_4, U_4') \geq 2\delta$.

Next, we need to show that $\cC_1 \cup \cC_2 \cup \cC_4$ is an $(n, 2\delta,\{k\})_q$ CDC. Using the same notations as in  Lemma \ref{Trivial}, it suffices to prove that $\dim(U_4 \cap K_1^{(t)}) \geq \delta$ and $\dim(U_4 \cap K_2^{(s)}) \geq \delta$  for all $t\in T_1, s \in T_2$ and for any codeword $U_4 \in \cC_4$.   It is clear that
    \begin{align*}
\dim(U_4+K_1^{(t)})&=\rank \left(\begin{array}{c:c}
        \begin{matrix}
        \bB_1 & \bN_1 & \bO_{a_1\times(k-T_2^{\min})}\\ 
        \tau_{\bB_1}(\bE_2) &\bN_2 &\bO_{a_2\times(k-T_2^{\min})} 
        \end{matrix} &
        \begin{matrix}
           \bO_{a_1\times(k-T_1^{\min})}& \bN_3 &\tau_{\bB_2}(\bE_3) \\
            \bO_{a_2\times(k-T_1^{\min})}&\bN_4 &\bB_2
        \end{matrix} \\
        \hdashline
        \bO_{(n_2-k+t) \times n_1} & \bO_{(n_2-k+t) \times (k-t)}~|~\bI_{n_2-k+t}
        \end{array}\right)\\
        &= a_1 + \rank(\bE_2|\bN_2) +(n_2 - k +t),
    \end{align*}
    where $\rank(\bE_2|\bN_2) \leq a_2 - \delta$.
Similarly, we have
\begin{align*}
\dim(U_4+K_2^{(s)})&=\rank \left(\begin{array}{c:c}
        \begin{matrix}
        \bB_1 & \bN_1 & \bO_{a_1\times(k-T_2^{\min})}\\ 
        \tau_{\bB_1}(\bE_2) &\bN_2 &\bO_{a_2\times(k-T_2^{\min})} 
        \end{matrix} &
        \begin{matrix}
           \bO_{a_1\times(k-T_1^{\min})}& \bN_3 &\tau_{\bB_2}(\bE_3) \\
            \bO_{a_2\times(k-T_1^{\min})}&\bN_4 &\bB_2
        \end{matrix} \\
        \hdashline
        \bI_{n_1-k+s}~|~\bO_{(n_2-k+s) \times (k-s)} & \bO_{(n_2-k+s) \times n_2}
        \end{array}\right)\\
        &= (n_1-k+s)+a_2+\rank(\bN_3|\bE_3),
    \end{align*}
    where $\rank(\bN_3|\bE_3) \leq a_1 -\delta.$ Then it is easy to check that
    \begin{align*}
         \dim(U_4 \cap K_1^{(t)}) &=\dim(U_4)+\dim(K_1^{(t)})-\dim(U_4 + K_1^{(t)}) \\
         &\geq k + (n_2-k+t) -[a_1+(a_2-\delta)+(n_2-k+t)]= \delta,\\
         \dim(U_4 \cap K_2^{(s)}) &= \dim(U_4)+\dim(K_2^{(s)})-\dim(U_4 + K_2^{(s)}) \\
         &\geq k+(n_1-k+s)-[(n_1-k+s)+a_2 + a_1 -\delta]=\delta.
    \end{align*}
    Finally, we prove that $\cC_3 \cup \cC_4$ is an $(n, 2\delta, \{k\})_q$ CDC. Suppose 
    \begin{align*}
    U_3=\rm{rs}
        \begin{pmatrix}
            \bA_1 & \bM_1 & \bO_{a_1 \times(2k-T_1^{\min}-T_2^{\min})}& \tau_{\bA_2}(\bF_3) & \bM_3\\
            \tau_{\bA_1}(\bF_2) & \bM_2 & \bO_{a_2 \times(2k-T_1^{\min}-T_2^{\min})}&
            \bA_2 & \bM_4
        \end{pmatrix}
    \end{align*}
    is a $k$-dimensional subspace in $\cC_3$.  By Equation (\ref{DS}), the subspace distance between $U_3$ and $U_4$ is 
    \begin{align*}
       d_S(U_3,U_4)&=2\rank 
       \begin{pmatrix}
            \bA_1 & \bM_1 & \bO_{a_1 \times(2k-T_1^{\min}-T_2^{\min})}& \tau_{\bA_2}(\bF_3) & \bM_3\\
            \tau_{\bA_1}(\bF_2) & \bM_2 & \bO_{a_2 \times(2k-T_1^{\min}-T_2^{\min})}&
            \bA_2 & \bM_4\\
            \bB_1 & \bN_1 & \bO_{a_1 \times (2k-T_1^{\min}-T_2^{\min})} & \bN_3 & \tau_{\bB_2}(\bE_3)\\
        \tau_{\bB_1}(\bE_2) & \bN_2 &\bO_{a_2 \times (2k-T_1^{\min}-T_2^{\min})} & \bN_4 & \bB_2
        \end{pmatrix} -2k\\
        &=2\rank 
       \begin{pmatrix}
            \bA_1 &\tau_{\bA_2}(\bF_3)&\bM_1  & \bM_3\\
            \tau_{\bA_1}(\bF_2) & \bA_2 &\bM_2 &
             \bM_4\\
            \bB_1 &\bN_3 & \bN_1  &  \tau_{\bB_2}(\bE_3)\\
        \tau_{\bB_1}(\bE_2) & \bN_4 &\bN_2  &  \bB_2
        \end{pmatrix} -2k
        =d_S(U_3'',U_4''),
    \end{align*}
    where
    \[U_3''=\rs \begin{pmatrix}
            \bA_1 &\tau_{\bA_2}(\bF_3)&\bM_1  & \bM_3\\
            \tau_{\bA_1}(\bF_2) & \bA_2 &\bM_2 &
             \bM_4
             \end{pmatrix} ~\text{and}~ \mathnormal{U_4''=} \rs
\begin{pmatrix}
    \bB_1 &\bN_3 & \bN_1  &  \tau_{\bB_2}(\bE_3)\\
        \tau_{\bB_1}(\bE_2) & \bN_4 &\bN_2  &  \bB_2
\end{pmatrix}.\]
 It is evident that $i(U_3'')$ has $k$ ones in the first $t_1+t_2-k+T_1^{\min}$ positions, whereas $i(U_4'')$ has at most $k-\delta$ ones in the same positions, since $a_1+\rank(\bE_2|\bN_4) \leq a_1+[(t_1-a_1)+(k-t_1-\delta)] = k-\delta$. By Lemma \ref{L-S-H},
\begin{align*}
    d_S(U_3,U_4)=d_S(U_3'',U_4'')\geq d_H(i(U_3''), i(U_4'')) \geq 2\delta.
\end{align*}
Therefore, $\bigcup_{i=1}^4 \cC_i$ is an $(n,\sum_{i=1}^4 |\cC_i|, 2\delta,\{k\})_q$ CDC.
\end{IEEEproof}

The construction of CDCs as outlined in Theorem \ref{Insert2} yields the following lower bounds for CDCs.

\begin{corollary}\label{C-Insert2}
Consider the same notations used in Theorem \ref{Insert2}. Then
    \begin{align*}
    A_q(n, 2\delta,\{k\}) &\geq \sum_{t \in T_1}\eta_t(\cX_1) \cdot \Delta(k,n_2+t-k,\delta)_q \\& +\sum_{s \in T_2}\eta_s(\cX_2) \cdot \Delta(k,n_1+s-k,\delta; T_1^{\min}-\delta-(k-s))_q  \\&+ A_q(t_1,2\delta, \{a_1\})\cdot \Delta(a_1, n_2-k+T_1^{\min}-a_2, \delta;a_1-\delta )_q  \\& \cdot \Delta(a_2, n_1-k+T_2^{\min}-a_1, \delta; a_2-\delta)_q \cdot A_q(t_2-k+T_1^{\min},2\delta, \{a_2\}) \\& \cdot f \cdot \Delta(a_1, n_1 -k+T_2^{\min}-t_1, \delta)_q \cdot \Delta(a_2, n_2-t_2, \delta)_q\\
        & +A_q(t_1,2\delta,\{a_1\}) \cdot A_q(n_2-t_2,2\delta,\{a_2\})
        \cdot 
        \Delta(a_2,n_1-k+T_2^{\min}-a_1,\delta;a_2-\delta)_q  \\
        &\cdot \Delta(a_1, n_1-k+T_2^{\min}-t_1, \delta)_q
        \cdot
      \Delta(a_1,n_2-k+T_1^{\min}-a_2,\delta;a_1-\delta)_q  \\ &\cdot \Delta(a_2,t_2-k+T_1^{\min},\delta; k-t_1-\delta)_q,
    \end{align*}
    where $f=\min\left\{\frac{\Delta(a_1,n_1-k+T_2^{\min}-t_1,b_1)_q}{\Delta(a_1,n_1-k+T_2^{\min}-t_1,\delta)_q},
        \frac{\Delta(a_2,n_2-t_2,b_2)_q}{\Delta(a_2,n_2-t_2,\delta)_q}
        \right\}$.
\end{corollary}
\begin{IEEEproof}
    By Lemma \ref{L-Subcode}, we can  construct $\cM_1^r$ and $\cM_4^r$ satisfying the conditions in Theorem \ref{Insert}, where $1 \leq r \leq f$. Consequently, the conclusion follows.
\end{IEEEproof}

\begin{example}\label{E-15-4-5}
    Set $n_1=5$, $n_2=10$, $\delta=2$, $k=5$, $T_1 = \{5\}$, and $T_2=\{5,4\}$ in Theorem \ref{Mix}. If we construct the $(10,4,3,\{5,4\})_q$ MDDC from Theorem \ref{T-ConstrMDDC}, then we have
    \begin{align*}
    |\cC_1 \cup \cC_2| &= q^{40}+\Delta(5,5,2;3)_q A_q(10,4,\{5\})+\Delta(5,4,2;2)_qN_q(10,2,5).
    \end{align*}
    Let $a_1=t_1=2$, $a_2=3$, $t_2=7$, and $b_1=b_2=1$. Note that $n_2-t_2 \geq a_2$ and $k-t_1\geq \delta$. From Corollary \ref{C-Insert2}, the cardinality of $\cC_3$ is 
    \begin{align*}
        |\cC_3|&=A_q(2,4,\{2\})  \cdot \Delta(2,7,2;0)_q\cdot \Delta(3,2,2;1)_q\cdot A_q(7,4,\{3\}) \cdot \Delta(2,2,2)_q \cdot \Delta(3,3,2)_q  \\ 
        & \cdot \min\left\{\frac{\Delta(2,2,1)_q}{\Delta(2,2,2)_q},  \frac{\Delta(3,3,1)_q}{\Delta(3,3,2)_q}
        \right\}\\
        &=1\cdot1\cdot1\cdot A_q(7,4,\{3\})\cdot q^2 \cdot q^6 \cdot q^2=A_q(7,4,\{3\})q^{10}.
    \end{align*}
    The cardinality of $\cC_4$ is 
    \begin{align*}
    |\cC_4| &= A_q(2,4,\{2\}) \cdot A_q(3,4,\{3\}) \cdot \Delta(3,2,2;1)_q \cdot \Delta(2,2,2)_q \cdot \Delta(2,7,2;0)_q \cdot \Delta(3,7,2;1)_q\\
    &=1\cdot1\cdot1\cdot q^2 \cdot 1 \cdot 1 = q^2.  
    \end{align*}
    For $q=2$, we have 
    \begin{align*}
   A_2(15,4,\{5\}) &\geq |\cC_1 \cup \cC_2| + |\cC_3| + |\cC_4|=1252457415410+340992+4\\&=1252457756406.
    \end{align*}
    This lower bound will be further improved by the multilevel construction in the next subsection.
\end{example}

\begin{theorem}\label{Insert2'}
Use the same notations as in Theorem \ref{Insert}. Assume that $n_2-t_2 \geq a_2$ and $k-t_1\geq 2\delta$. Let $c_1$ and $c_2$ be positive integers such that $c_1 + c_2 \geq \delta$, and $1 \leq c_i \leq \delta$ for $i=1,2$. Let $\hcB_1 = \{\hbB_1 \in 
\BF_q^{a_1 \times t_1}: \rank(\hbB_1) = a_1, \hbB_1~\text{in}~\RREF\}$ and $\hcB_2 = \{\hbB_2 \in \BF_q^{a_2 \times (n_2-t_2)}: \rank(\hbB_2)=a_2, \hbB_2~\text{in}~\RREF\}$ be SC-representations of two $(t_1, 2\delta, \{a_1\})_q$ and $(n_2-t_2, 2\delta, \{a_2\})_q$ CDCs, respectively.  Let 
 \begin{align*}
     \hcN_2=\left\{ (\hbE_2~| ~\hbN_2): \hbE_2 \in \BF_q^{a_2 \times (t_1-a_1)},\hbN_2 \in \BF_q^{a_2 \times (n_1 - k + T_2^{\min}-t_1)} 
\right\} 
\end{align*}
and
 \begin{align*}
     \hcN_3=\left\{(\hbN_3~|~\hbE_3): \hbN_3 \in \BF_q^{a_1 \times (t_2-k+T_1^{\min})}, \hbE_3 \in \BF_q^{a_1 \times (n_2-t_2-a_2)} \right\}
 \end{align*}
be two $(a_2 \times (n_1 -k + T_2^{\min} -a_1), \delta; a_2-\delta)_q$ and $(a_1 \times (n_2-k+T_1^{min}-a_2), \delta; a_1-\delta)_q$ RRMCs. Given another integer l,
let $\hcN_1^r$ be an $(a_1 \times (n_1 - k + T_2^{\min}-t_1), 
 \delta)_q$ MRD code and $\hcN_4^r$ be an $(a_2 \times (t_2-k+T_1^{\min}), \delta; k - t_1 - \delta)_q$ RRMC for all $1 \leq r \leq l$. Assume that $\hbN \neq \hbN'$ and $\rank(\hbN-\hbN') \geq c_1$ for $\hbN \in \hcN_1^r$ and  $\hbN' \in \hcN_1^{r'}$ for all $1 \leq r < r' \leq l$.
 Assume that $\hbN \neq \hbN'$ and $\rank(\hbN-\hbN') \geq c_2$ for $\hbN \in \hcN_4^r$ and $\hbN' \in \hcN_4^{r'}$ for all $1 \leq r < r' \leq l$.
Define $\hcC_4 =\bigcup_{r=1}^l \hcC_4^{(r)}$,
\begin{align*}
    \hcC_4^{(r)} = 
    \left\{\rs
    \begin{pmatrix}
        \hbB_1 & \hbN_1 & \bO_{a_1 \times (2k-T_1^{\min}-T_2^{\min})} & \hbN_3 & \tau_{\hbB_2}(\hbE_3)\\
        \tau_{\hbB_1}(\hbE_2) & \hbN_2 &\bO_{a_1 \times (2k-T_1^{\min}-T_2^{\min})} & \hbN_4 & \hbB_2
    \end{pmatrix}
    \right\},
\end{align*}
where $\hbB_i \in \hcB_i$ for $i = 1,2$, $(\hbE_2|\hbN_2) \in \hcN_2$, $(\hbN_3 |\hbE_3) \in \hcN_3$, and $\hbN_i \in \hcN_i^r$ for $i=1,4$.
Then $\bigcup_{i=1}^3 \cC_i \cup \hcC_4$ is an $(n, \sum_{i=1}^3|\cC_i|+|\hcC_4|, 2\delta, \{k\})_q$ CDC.    
\end{theorem}
\begin{IEEEproof}
    First, we need to show that $\hcC_4$ is an $(n,2\delta,\{k\})_q$ CDC.
    It is easy to verify  that the elements of $\hcC_4$ are $k$-dimensional subspaces of $\BF_q^n$.
    Let
\begin{align*}
    \hU_4 &= \rs(\hbG_1) =
    \rs
    \begin{pmatrix}
        \hbB_1 & \hbN_1 & \bO_{a_1 \times (2k-T_1^{\min}-T_2^{\min})} & \hbN_3 & \tau_{\hbB_2}(\hbE_3)\\
        \tau_{\hbB_1}(\hbE_2) & \hbN_2 &\bO_{a_2 \times (2k-T_1^{\min}-T_2^{\min})} & \hbN_4 & \hbB_2
    \end{pmatrix},\\
    \hU_4' &= \rs(\hbG_2) =
    \rs
    \begin{pmatrix}
        \hbB'_1 & \hbN'_1 & \bO_{a_1 \times (2k-T_1^{\min}-T_2^{\min})} & \hbN'_3 & \tau_{\hbB'_2}(\hbE'_3)\\
        \tau_{\hbB'_1}(\hbE'_2) & \hbN'_2 &\bO_{a_2 \times (2k-T_1^{\min}-T_2^{\min})} & \hbN'_4 & \hbB'_2
    \end{pmatrix}
\end{align*}
be two distinct codewords in $\hcC_4$. The intersection of $\hU_4$ and $\hU_4'$ is 
\begin{align*}
    \hU_4 \cap \hU_4' = \left\{
    (\alpha_1, \alpha_2)\hbG_1=(\beta_1,\beta_2)\hbG_2: \alpha_i,\beta_i \in \BF_q^{a_i},i=1,2
    \right\}.
\end{align*}

We analyze the dimension of $\hU_4 \cap \hU_4'$ in three cases.
\begin{itemize}
 \item [(1)] If $\hbB_1 \neq \hbB'_1$, then $\dim(\hU_4 \cap \hU_4') \leq \dim(\rs(\hbB_1) \cap \rs(\hbB'_1)) + a_2 \leq (a_1-\delta)+a_2=k-\delta$.
    \item[(2)] If $\hbB_1 = \hbB'_1$ and $\hbB_2 \neq \hbB'_2$, then $\dim(\hU_4 \cap \hU_4') \leq \dim(\rs(\hbB_2) \cap\rs(\hbB'_2))+a_1 \leq (a_2 - \delta) +a_1=k-\delta$. 
    \item[(3)]If $\hbB_1 = \hbB'_1$ and $\hbB_2 = \hbB'_2$, we obtain that
    \begin{align*}
        \begin{cases}
            \alpha_1 \hbB_1+\alpha_2 \tau_{\hbB_1}(\hbE_2) =\beta_1 \hbB_1+\beta_2 \tau_{\hbB_1}(\hbE'_2)
            \\
            \alpha_1\tau_{\hbB_2}(\hbE_3)+\alpha_2\hbB_2=\beta_1\tau_{\hbB_2}(\hbE'_3)+\beta_2\hbB_2
        \end{cases}.
        \end{align*}
        According to the definitions of $\tau_{\hbB_1}(\hbE_2), \tau_{\hbB_1}(\hbE'_2)$, $\tau_{\hbB_2}(\hbE_3),$ and $ \tau_{\hbB_2}(\hbE'_3)$, we can deduce that $\alpha_1 = \beta_1$ and $\alpha_2 = \beta_2$. Hence,
        \begin{align*}
            \dim(\hU_4 \cap \hU_4')=\dim(\{(\alpha_1, \alpha_2): (\alpha_1, \alpha_2)\hbG 
            =\mathbf{0}, \alpha_i \in \BF_q^{a_i}, i=1,2
            \}),
        \end{align*}
        where $\hbG=\begin{pmatrix}
            \hbG_{11} & \hbG_{12}\\
            \hbG_{21} & 
            \hbG_{22}
        \end{pmatrix}$ with $\hbG_{11} = (\bO  | \hbN_1-\hbN'_1 )$, $\hbG_{12} = (\hbN_3|\hbE_3)-(\hbN'_3|\hbE'_3)$,$\hbG_{21} =(\hbE_2|\hbN_2)-(\hbE'_2|\hbN'_2)$, and $\bG_{22}=(\hbN_4 -\hbN'_4|
                \bO)$.
                
We analyze the following subcases.

\begin{itemize}
    \item If $\hbG_{12} \neq \bO$ or $\hbG_{21} \neq \bO$, then
    \begin{align*}
    \dim(\hU_4 \cap \hU_4')&\leq k-\max\{\rank((\hbN_3|\hbE_3)-(\hbN'_3|\hbE'_3)), \rank((\hbE_2|\hbN_2)-(\hbE'_2|\hbN'_2))\} \\ &\leq k-\delta.
    \end{align*}
    \item When $\hbG_{12} = \bO$ and $\hbG_{21} = \bO$, then 
    \begin{align*}
     \dim(\hU_4 \cap \hU_4')&=
    \dim(\{\alpha_1 \in \BF_q^{a_1}:\alpha_1 \hbG_{11}=\mathbf{0}\}) + \dim(\{\alpha_2 \in \BF_q^{a_2}:\alpha_2 \hbG_{22}=\mathbf{0}\}) \\&=a_1-\rank(\bO|\hbN_1-\hbN'_1)+a_2-\rank(\hbN_4-\hbN'_4|\bO)\\
     &=k -\rank(\hbN_1-\hbN'_1) -\rank(\hbN_4-\hbN'_4).
  \end{align*}
  \begin{itemize}
         \item[(\romannumeral1)] If $r =r'$, then $\hbN_1, \hbN'_1 \in \hcN_1^r$ and $\hbN_4, \hbN'_4 \in \hcN_4^r$. If $\hbN_1 \neq \hbN'_1$, then $\dim(\hU_4 \cap \hU_4') \leq k -\rank(\hbN_1-\hbN'_1) \leq k- \delta$. If $\hbN_1 = \hbN'_1$, we have $\hbN_4 \neq \hbN'_4$, and then $\dim(\hU_4 \cap \hU_4') = k - \rank(\hbN_4 - \hbN'_4) \leq k-\delta.$
         \item[(\romannumeral2)] If $r \neq r'$, then 
         \begin{align*}
           \dim(\hU_4 \cap \hU_4')&= k -\rank(\hbN_1-\hbN'_1) -\rank(\hbN_4-\hbN'_4)\\
           &\leq k - b_1 -b_2 \leq k-\delta.
         \end{align*}
     \end{itemize}
\end{itemize}
\end{itemize}

As a result, $\dim(\hU_4 \cap \hU_4') \leq k-\delta$. It follows that $d_S(\hU_4, \hU_4') \geq 2\delta$.

Next, we need to prove that $\cC_1 \cup \cC_2 \cup \hcC_4$ is an $(n, 2\delta,\{k\})_q$ CDC. With the same notations used in Lemma \ref{Trivial}, it suffices to prove that $\dim(\hU_4 \cap K_1^{(t)}) \geq \delta$ and $\dim(\hU_4 \cap K_2^{(s)}) \geq \delta$ for all $t \in T_1$, $s \in T_2$ and codeword $\hU_4 \in \hcC_4$. It is obvious that
\begin{align*}
\dim(\hU_4+K_1^{(t)})&=\rank \left(\begin{array}{c:c}
        \begin{matrix}
        \hbB_1 & \hbN_1 & \bO_{a_1\times(k-T_2^{\min})}\\ 
        \tau_{\hbB_1}(\hbE_2) &\hbN_2 &\bO_{a_2\times(k-T_2^{\min})} 
        \end{matrix} &
        \begin{matrix}
           \bO_{a_1\times(k-T_1^{\min})}& \hbN_3 &\tau_{\hbB_2}(\hbE_3) \\
            \bO_{a_2\times(k-T_1^{\min})}&\hbN_4 &\hbB_2
        \end{matrix} \\
        \hdashline
        \bO_{(n_2-k+t) \times n_1} & \bO_{(n_2-k+t) \times (k-t)}~|~\bI_{n_2-k+t}
        \end{array}\right)\\
        &= a_1 + \rank(\hbE_2|\hbN_2) +(n_2 - k +t),
    \end{align*}
    where $\rank(\hbE_2|\hbN_2) \leq a_2-\delta$.
    Similarly, we obtain that
    \begin{align*}
\dim(\hU_4+K_2^{(s)})&=\rank \left(\begin{array}{c:c}
        \begin{matrix}
        \hbB_1 & \hbN_1 & \bO_{a_1\times(k-T_2^{\min})}\\ 
        \tau_{\hbB_1}(\hbE_2) &\hbN_2 &\bO_{a_2\times(k-T_2^{\min})} 
        \end{matrix} &
        \begin{matrix}
           \bO_{a_1\times(k-T_1^{\min})}& \hbN_3 &\tau_{\hbB_2}(\hbE_3) \\
            \bO_{a_2\times(k-T_1^{\min})}&\hbN_4 &\hbB_2
        \end{matrix} \\
        \hdashline
        \bI_{n_1-k+s}~|~\bO_{(n_1-k+s) \times (k-s)}&\bO_{(n_1-k+s) \times n_2}
   \end{array}\right)\\
        &= a_2 + \rank(\hbN_3|\hbE_3) +(n_1 - k +s),
    \end{align*}
    where $\rank(\hbN_3|\hbE_3) \leq a_1 - \delta$. As a result,
    \begin{align*}
         \dim(\hU_4 \cap K_1^{(t)}) &=\dim(\hU_4)+\dim(K_1^{(t)})-\dim(\hU_4 + K_1^{(t)}) \\
         &\geq k + (n_2-k+t) -[a_1+(a_2-\delta)+(n_2-k+t)]= \delta,\\
          \dim(\hU_4 \cap K_2^{(s)}) &=\dim(\hU_4)+\dim(K_2^{(s)})-\dim(\hU_4 + K_2^{(s)}) \\
         &\geq k + (n_1-k+s) -[a_2+(a_1-\delta)+(n_1-k+s)]= \delta.
    \end{align*}
    
    It follows that $\cC_1 \cup \cC_2 \cup \hcC_4$ is an $(n, 2\delta, \{k\})_q$ CDC.
    
Finally, we show that $\cC_3 \cup \hcC_4$ is an $(n,2\delta,\{k\})_q$ CDC. Let
\begin{align*}
   U_3 = \rm{rs}
        \begin{pmatrix}
            \bA_1 & \bM_1 & \bO_{a_1 \times(2k-T_1^{\min}-T_2^{\min})}& \tau_{\bA_2}(\bF_3) & \bM_3\\
            \tau_{\bA_1}(\bF_2) & \bM_2 & \bO_{a_2 \times(2k-T_1^{\min}-T_2^{\min})}&
            \bA_2 & \bM_4
        \end{pmatrix}
\end{align*}
be a codeword in $\cC_3$. By Equation (\ref{DS}), the subspace distance between $U_3$ and $\hU_4$ is 
\begin{align*}
    d_S(U_3,\hU_4)
    &=2\rank 
       \begin{pmatrix}
            \bA_1 & \bM_1 & \bO_{a_1 \times(2k-T_1^{\min}-T_2^{\min})}& \tau_{\bA_2}(\bF_3) & \bM_3\\
            \tau_{\bA_1}(\bF_2) & \bM_2 & \bO_{a_2 \times(2k-T_1^{\min}-T_2^{\min})}&
            \bA_2 & \bM_4\\
            \hbB_1 & \hbN_1 & \bO_{a_1 \times (2k-T_1^{\min}-T_2^{\min})} & \hbN_3 & \tau_{\hbB_2}(\hbE_3)\\
        \tau_{\hbB_1}(\hbE_2) & \hbN_2 &\bO_{a_2 \times (2k-T_1^{\min}-T_2^{\min})} & \hbN_4 & \hbB_2
        \end{pmatrix} -2k\\
        &=2\rank 
       \begin{pmatrix}
            \bA_1 &\tau_{\bA_2}(\bF_3)&\bM_1  & \bM_3\\
            \tau_{\bA_1}(\bF_2) & \bA_2 &\bM_2 &
             \bM_4\\
            \hbB_1 &\hbN_3 & \hbN_1  &  \tau_{\hbB_2}(\hbE_3)\\
        \tau_{\hbB_1}(\hbE_2) & \hbN_4 &\hbN_2  &  \hbB_2
        \end{pmatrix} -2k
        =d_S(U_3'',\hU_4''),
\end{align*}
where
\[U_3''=\rs \begin{pmatrix}
            \bA_1 &\tau_{\bA_2}(\bF_3)&\bM_1  & \bM_3\\
            \tau_{\bA_1}(\bF_2) & \bA_2 &\bM_2 &
             \bM_4
             \end{pmatrix} ~\text{and}~ \mathnormal{\hU_4''}= \rs\mathnormal{
\begin{pmatrix}
    \hbB_1 &\hbN_3 & \hbN_1  &  \tau_{\hbB_2}(\hbE_3)\\
     \tau_{\hbB_1}(\hbE_2) & \hbN_4 &\hbN_2  &  \hbB_2
\end{pmatrix}}.\]
 It is obvious that $i(U_3'')$ has $k$ ones in the first $t_1+t_2-k+T_1^{\min}$ positions. But $i(U_4'')$ has at most $k-\delta$ ones in the first $t_1+t_2-k+T_1^{\min}$ positions since $a_1+\rank(\hbE_2|\hbN_4) \leq a_1+[(t_1-a_1)+(k-t_1-\delta)]=k-\delta$. By Lemma \ref{L-S-H},
\[d_S(U_3,\hU_4)=d_S(U_3'',\hU_4'')\geq d_H(i(U_3''), i(\hU_4'')) \geq  2\delta.
\]

To summarize, $\bigcup_{i=1}^3 \cC_i \cup \hcC_4$
is an $(n,\sum_{i=1}^3 |\cC_i|+|\hcC_4|, 2\delta, \{k\})_q$ CDC.
\end{IEEEproof}

From the construction of CDC in Theorem \ref{Insert2'}, we can derive the following lower bounds for CDCs.
 
\begin{corollary}\label{C-Insert2'}
    Consider the same notations used in Theorem \ref{Insert2'}. Then
    \begin{align*}
        A_q(n,2\delta,\{k\}) &\geq \sum_{t \in T_1}\eta_t(\cX_1) \cdot \Delta(k,n_2+t-k,\delta)_q \\& +\sum_{s \in T_2}\eta_s(\cX_2) \cdot \Delta(k,n_1+s-k,\delta; T_1^{\min}-\delta-(k-s))_q  \\&+A_q(t_1,2\delta, \{a_1\})\cdot \Delta(a_1, n_2-k+T_1^{\min}-a_2, \delta;a_1-\delta )_q  \\&\cdot \Delta(a_2, n_1-k+T_2^{\min}-a_1, \delta; a_2-\delta)_q\cdot A_q(t_2-k+T_1^{\min},2\delta, \{a_2\}) \\&\cdot f\cdot  \Delta(a_1, n_1 -k+T_2^{\min}-t_1, \delta)_q \cdot \Delta(a_2, n_2-t_2, \delta)_q\\
        &+ A_q(t_1, 2\delta,\{a_1\})\cdot A_q(n_2-t_2,2\delta,\{a_2\}) \cdot \Delta(a_2 ,n_1-k+T_2^{\min}-a_1, \delta;a_2-\delta)_q \\&\cdot 
    \Delta(a_1 ,n_2-k+T_1^{\min}-a_2, \delta; a_1-\delta)_q \cdot \Delta(a_1, n_1-k+T_2^{\min}-t_1, \delta)_q\\
    &\cdot[l \cdot \Delta(a_2,t_2-k+T_1^{\min},\delta;k-t_1-\delta)_q-(l-1)],\end{align*}
    where $f=\min\left\{\frac{\Delta(a_1,n_1-k+T_2^{\min}-t_1,b_1)_q}{\Delta(a_1,n_1-k+T_2^{\min}-t_1,\delta)_q},
        \frac{\Delta(a_2,n_2-t_2,b_2)_q}{\Delta(a_2,n_2-t_2,\delta)_q}
        \right\}$ and \\$l=\min \left\{\frac{\Delta(a_1,n_1-k+T_2^{\min}-t_1, c_1)_q}{\Delta(a_1,n_1-k+T_2^{\min}-t_1, \delta)_q}, \frac{\Delta(a_2 , t_2-k+T_1^{\min}, c_2)_q}{\Delta(a_2 ,t_2-k+T_1^{\min}, \delta)_q}\right\}$.
\end{corollary}

\begin{IEEEproof}
By Lemma \ref{L-Subcode}, we can construct $\hcN_1^r$ that satisfies the conditions in Theorem \ref{Insert2'}, where $r \in \left\{1,2,\dots, \frac{\Delta(a_1,n_1-k+T_2^{\min}-t_1, c_1)_q}{\Delta(a_1,n_1-k+T_2^{\min}-t_1, \delta)_q}\right\}$. By Lemma \ref{L-Subcode}, there exists an $[a_2 \times (t_2-k+T_1^{\min}), c_2]_q$ MRD code that contains the unique  $[a_2 \times (t_2-k+T_1^{\min}), \delta]_q$ MRD code $\hcN_4^{(1)}$ as well as  other nonlinear MRD codes $\hcN_4^{(r)}$ with $r \in \left\{2,3, \dots,\frac{\Delta(a_2 ,t_2-k+T_1^{\min}, c_2)_q}{\Delta(a_2 ,t_2-k+T_1^{\min}, \delta)_q}\right\}$.
Let $\hcN_4^{r} = \{\hbN \in \hcN_4^{(r)}: \rank(\hbN) \leq k - t_1 - \delta\}$ for $r \in \left\{1,2, \dots,\frac{\Delta(a_2 ,t_2-k+T_1^{\min}, c_2)_q}{\Delta(a_2 ,t_2-k+T_1^{\min}, \delta)_q}\right\}$. 

Set $l=\min \left\{\frac{\Delta(a_1,n_1-k+T_2^{\min}-t_1, c_1)_q}{\Delta(a_1,n_1-k+T_2^{\min}-t_1, \delta)_q}, \frac{\Delta(a_2 ,t_2-k+T_1^{\min}, c_2)_q}{\Delta(a_2,t_2-k+T_1^{\min}, \delta)_q}\right\}$. 
Then $|\hcC_4|=|\hcB_1|\cdot|\hcB_2|\cdot|\hcN_2|\cdot|\hcN_3|\cdot|\hcN_1^1| \cdot (|\hcN_4^1|+(l-1)|\hcN_4^2|)$, where $|\hcN_4^2|=|\hcN_4^1|-1$ and 
$|\hcN_4^1|=\Delta(a_2,t_2-k+T_1^{\min},\delta;k-t_1-\delta)_q$.  The desired conclusion follows.
\end{IEEEproof}

\begin{example}\label{E-18-4-6}
    Set $n_1=6$, $n_2=12$, $\delta = 2$, $k=6$, $T_1=\{6\}$, and $T_2=\{5, 6\}$ in Theorem \ref{Mix}. If we construct the $(12,4,3, \{5,6\})_q$ MDDC from Theorem \ref{T-ConstrMDDC}, then we have
    \begin{align*}
       |\cC_1 \cup \cC_2| = q^{60}+\Delta(6,6,2;4)_q A_q(12,4,\{6\})+\Delta(6,5,2;3)_qN_q(12,2,6).
    \end{align*}
    Let $a_1 =t_1=2$, $a_2=4$, $t_2=8$, and $b_1=b_2=1$. By Theorem \ref{Insert}, we have
    \[
    f=\min\left\{\frac{\Delta(2,3,1)_q}{\Delta(2,3,2)_q}, \frac{\Delta(4,4,1)_q}{\Delta(4,4,2)_q}\right\}=q^3,\] and the cardinality of $\cC_3$ is
    \begin{align*}
    |\cC_3|&=A_q(2,4,\{2\})\cdot\Delta(2,8,2;0)_q \cdot \Delta(4,3,2;2)_q \cdot A_q(8,4,\{4\}) \cdot f \cdot \Delta(2,3,2)_q \cdot \Delta(4,4,2)_q\\
    &=1\cdot 1 \cdot (q^6 + q^5 + q^4 - q^2 - q)\cdot A_q(8,4,\{4\}) \cdot q^3 \cdot q^3 \cdot q^{12}\\& =A_q(8,4,\{4\}) q^{18}(q^6 + q^5 + q^4 - q^2 - q).
    \end{align*}
    
    Notice that $n_2-t_2 \geq a_2$ and $k-t_1 \geq 2\delta$. Let $c_1=c_2=1$. By Theorem \ref{Insert2'}, we have
    \[l=\min\left\{\frac{\Delta(2,3,1)_q}{\Delta(2,3,2)_q},\frac{\Delta(4,8,1)_q}{\Delta(4,8,2)_q}\right\}=q^3,\]
    and the cardinality of $\hcC_4$ is
    \begin{align*}
        |\hcC_4|&=A_q(2,4,\{2\}) \cdot A_q(4,4,\{4\}) \cdot \Delta(4,3,2;2)_q\cdot \Delta(2,8,2;0)_q \cdot \Delta(2,3,2)_q\\&\cdot(l \cdot \Delta(4,8,2;2)_q-(l-1))\\& =1 \cdot 1\cdot (q^6 + q^5 + q^4 - q^2 - q )\cdot 1 \cdot q^3 \cdot[q^3 \cdot (q^{12} + q^{11} + 2q^{10}  \\&+q^9 + q^8 - q^4 - q^3 - 2q^2 - q)-(q^3-1)]\\
        &= q^{24} + 2q^{23} + 4q^{22} + 4q^{21} + 3q^{20} - 2q^{18} - 3q^{17} - 3q^{16} - 3q^{15} \\&- 4q^{14} - 4q^{13} - 3q^{12} + 2q^{10} + 4q^9 + 3q^8 + 2q^7 - q^5 - q^4.
    \end{align*}

    By Theorem \ref{Insert2},  
    the cardinality of $\cC_4$ is
    \begin{align*}
        |\cC_4|&=A_q(2,4,\{2\})\cdot A_q(4,4,\{4\}) \cdot \Delta(4,3,2;2)_q \cdot \Delta(2,3,2)_q \cdot \Delta(2,8,2;0)_q \cdot \Delta(4,8,2;2)_q\\
        &=1\cdot 1 \cdot (q^6 + q^5 + q^4 - q^2 - q)\cdot q^3 \cdot 1 \cdot (q^{12} + q^{11} + 2q^{10} + q^9 + q^8 - q^4 - q^3 - 2q^2 - q)\\ &= q^{21} + 2q^{20} + 4q^{19} + 4q^{18} + 3q^{17} - 2q^{15} - 3q^{14} - 3q^{13} - 3q^{12} - 4q^{11} - 4q^{10} \\&- 2q^9 + q^8 + 3q^7 + 3q^6 + q^5.
    \end{align*}
   For $q=2$, we have $|\cC_1 \cup \cC_2|=1321068515713406208$, $|\cC_3|= 133406654464$, $|\hcC_4| = 60548048$, and $|\cC_4|= 7569248$. Obviously, when $k-t_1 \geq 2\delta$, the method presented in Theorem \ref{Insert2'} is  superior to the construction in Theorem \ref{Insert2}.
\end{example}

\subsection{Multilevel Construction}
Indeed, the mixed dimension construction and the inserting constructions mentioned above can be further enhanced using the multilevel construction\cite{Multilevel}. 

The subsequent theorem combines the multilevel construction from lifting RFDRMCs with the mixed dimension construction. For a vector $\bw=(w_1 w_2 \cdots w_n) \in \BF_2^n$,  define $\bw^{[i,j]}=(w_i \cdots w_j)$ as the substring of $\bw$ that starts at the $i$-th coordinate and ends at the $j$-th coordinate, where $1 \leq i \leq j \leq n$.

\begin{theorem}\label{M-Insert}
Take the same notations as in Theorem \ref{Mix}. Let $\cS \subseteq \BF_2^n$ be a set of identifying vectors $\bw_i$ with Hamming weight $k$ such that $\wt(\bw_i^{[1,n_1+T_2^{\min}-k]}) \geq \delta$, $\wt(\bw_i^{[n_1+k-T_1^{\min}+1,n]}) \geq \delta$, and $d_H(\bw_i, \bw_j) \geq 2\delta$ for all $\bw_i, \bw_j \in \cS$, where $i, j \in \{1,2,\dots,|\cS|\}$ and $i \neq j$.
For each $\bw_i \in \cS$, let $\cF_i$ denote the Ferrers diagram of $EF(\bw_i)$, and let $r_i$ denote the Hamming weight of $\bw_i^{[1,n_1+T_2^{\min}-k]}$. If there exists 
an $(\cF_i,\delta;r_i, n_2 - T_2^{\min}+r_i, r_i-\delta)_q$ RFDRMC $\cC_{\cF_i}$ for all $\bw_i \in \cS$, then $\tcC_1 = \bigcup_{i=1}^{|\cS|} \sL (\cC_{\cF_i})$ is an $(n, 2\delta, \{k\})_q$ CDC. Furthermore, $\bigcup_{i=1}^2 \cC_i \cup  \tcC_1$ is also an $(n, \sum_{i=1}^2|\cC_i|+|\tcC_1|, 2\delta, \{k\})_q$ CDC.
\end{theorem}
\begin{IEEEproof}
 According to Theorem \ref{T-LFRDMC},  $\tcC_1$ is an $(n,2\delta, \{k\})_q$ CDC. It remains to determine the distance among $\cC_1, \cC_2$, and $\tcC_1$.

For $U_1 \in \cC_1$ and $W \in \tcC_1$, let $\bu_1$ and $\bw$ be their corresponding identifying vectors. It is obvious that $\bu_1$ has $k$ ones in the first $n_1+k-T_1^{\min}$ positions. By Lemma \ref{L-S-H}, we have
\begin{align*}
d_S(U_1, W) &\geq d_H(\bu_1, \bw) \\&= d_H(\bu_1^{[1,n_1+k-T_1^{\min}]}, \bw^{[1,n_1+k-T_1^{\min}]})+d_H(\bu_1^{[n_1+k-T_1^{\min}+1,n]}, \bw^{[n_1+k-T_1^{\min}+1,n]})\\ & \geq [k-(k-\delta)] + \delta = 2\delta
.\end{align*}

For $U_2 \in \cC_2$, there exists a matrix 
$\bU_2^1 \in \BF_q^{k \times (n_1+T_2^{\min}-k)}$ and a full rank matrix $\bU_2^2 \in \BF_q^{k \times (n_2+k-T_2^{\min})}$ such that $U_2 = \rs(\bU_2^1~\bU_2^2)$. For a subspace $W \in \tcC_1$, let $\xi(W) = (\bW_1~ \bW_2)$, where $\bW_1 \in \BF_q^{k \times (n_1+T_2^{\min}-k)}$ and $\bW_2 \in \BF_q^{k \times (n_2+k-T_2^{\min})}$. By Equation (\ref{DS}), we have
    \begin{align*}
        d_S(U_2, W) =
        2\cdot \rank\begin{pmatrix}
            \bU_2^1 & \bU_2^2\\
        \bW_1 & \bW_2
        \end{pmatrix}-2k = 2\cdot \rank\begin{pmatrix}
            \bU_2^2 & \bU_2^1\\
        \bW_2 & \bW_1
        \end{pmatrix}-2k=d_S(U_2',W'),
    \end{align*}
    where $U_2'=\rs(\bU_2^2~\bU_2^1)$ and $W'=\rs(\bW_2~\bW_1)$.

      If $\rank(\bW_2) \leq k-\delta$, it is easy to verify that $d_S(U_2', W') \geq d_H(i(U_2'), i(W')) \geq [k-(k-\delta)] +\delta = 2\delta$. It suffices to determine the value of $\rank(\bW_2)$. Let $r=\wt(\bw^{[1,n_1+T_2^{\min} - k]})$. We partition $\xi(W)=(\bW_1~\bW_2)$ into  four blocks 
    \[\begin{pmatrix}
        \bG_1 & \bG_2\\
        \bG_3 & \bG_4
    \end{pmatrix},\] 
    where $\bG_1 \in \BF_q^{r \times (n_1 + T_2^{\min}-k)}$, $\bG_2 \in \BF_q^{r \times (n_2 + k- T_2^{\min})}$, $\bG_3 \in \BF_q^{(k-r)\times (n_1 + T_2^{\min}-k)}$, and $\bG_4 \in \BF_q^{(k-r) \times (n_2 + k- T_2^{\min})}$. Assume the Ferrer diagram of $W$ is
 \[\cF=\begin{pmatrix}
     \cF_1 & \cF_2\\
      & \cF_3
 \end{pmatrix},\]
 where  $\cF_2$ is an $r \times (n_2- T_2^{\min}+r)$ Ferrers diagram,  and the size of $\cF_i$ depends on $\bw$ for $i=1,3$. Note that $\cF_2$ is full  since $r$ pivot elements are in $\bG_1$ and the other $k-r$ pivot elements are in $\bG_4$.  Then there exists a codeword $\bX \in \cC_{\cF}$ such that $\cF(W) = \bX_{\cF}$ and $\sigma(\bX)_{r,n_2-T_2^{\min}+r}$ is embedded in $\bG_2$. Let the columns of $\bW_2$ be indexed by $\{1,2,\dots,n_2+k-T_2^{\min}\}$. Denote the index set of the $k-r$ pivot columns in $\bW_2$ by $\Lambda_1$, and the complement set $\{1,2,\dots,n_2+k-T_2^{\min}\}\backslash \Lambda_1$ by $\Lambda_2$. It implies that
    \begin{align*}
        \rank(\bW_2) &=\rank
        \begin{pmatrix}
            \bG_2\\
            \bG_4
        \end{pmatrix}=
       \rank \begin{pmatrix}
    \bG_2^{\Lambda_1} & \bG_2^{\Lambda_2}\\
    \bG_4^{\Lambda_1} & \bG_4^{\Lambda_2}
    \end{pmatrix}=
   \rank \begin{pmatrix}
        \bO_{r \times (k-r)} & \bG_2^{\Lambda_2}\\
        \bI_{k-r} & \bG_4^{\Lambda_2}
    \end{pmatrix}
    \\&=
    \rank(\sigma(\bX)_{r , n_2-T_2^{\min}+r}) +(k-r)\\& \leq (r-\delta)+(k-r)=k-\delta,        \end{align*}
where $\bG_l^{\Lambda_j}$ denotes the projection of $\bG_l$ onto the columns indexed by $\Lambda_j$, for $l=2, 4$ and $j=1, 2$.

In summary, $\bigcup_{i=1}^2 \cC_i \cup  \tcC_1$ is an $(n, \sum_{i=1}^2|\cC_i|+|\tcC_1|, 2\delta, \{k\})_q$ CDC.
\end{IEEEproof}

Applying the same technique, the next two theorems introduce two improved constructions
based on the inserting constructions from Theorems \ref{Insert2} and \ref{Insert2'}.

\begin{theorem}\label{M-Insert2}
    Following the same notations as in Theorem \ref{Insert2}, let $\cS \subseteq \BF_2^n$ be a set of identifying vectors $\bw_i$ with Hamming weight $k$ such that $\wt(\bw_i^{[1,n_1+T_2^{\min}-k]}) \geq \delta$, $\wt(\bw_i^{[n_1+k-T_1^{\min}+1,n]}) \geq \delta$, $\wt(\bw_i^{[1,t_1]})=0$, and $d_H(\bw_i, \bw_j) \geq 2\delta$ for all $\bw_i, \bw_j \in \cS$, where $i, j \in \{1,2,\dots,|\cS|\}$ and $i \neq j$.
For each $\bw_i \in \cS$, let $\cF_i$ denote the Ferrers diagram of $EF(\bw_i)$, and let $r_i$ denote the Hamming weight of $\bw_i^{[1,n_1+T_2^{\min}-k]}$. If there exists 
an $(\cF_i,\delta;r_i, n_2 - T_2^{\min}+r_i, r_i-\delta)_q$ RFDRMC $\cC_{\cF_i}$ for all $\bw_i \in \cS$, then $\tcC_2 = \bigcup_{i=1}^{|\cS| } \sL (\cC_{\cF_i})$ is an $(n, 2\delta, \{k\})_q$ CDC. Furthermore, $\bigcup_{i=1}^4 \cC_i \cup  \tcC_2$ is also an $(n, \sum_{i=1}^4|\cC_i|+|\tcC_2|, 2\delta, \{k\})_q$ CDC.
\end{theorem}

\begin{IEEEproof}
    According to Theorem \ref{M-Insert}, $\tcC_2$ is a special case of $\tcC_1$. Then $\bigcup_{i=1}^2 \cC_i \cup \tcC_2$ is an $(n, \sum_{i=1}^2|\cC_i|+|\tcC_2|, 2\delta, \{k\})_q$ CDC. 
    
    For subspaces $U_3 \in \cC_3$, $U_4 \in \cC_4$,  and $W \in \tcC_2$, it is obvious 
 that $i(U_3)$ and $i(U_4)$  have at least $a_1$ ones in the first $t_1$ positions, whereas $i(W)$ has all zeros in the first $t_1$ positions. Since $a_1\geq \delta$, for $j=3,4$, Lemma \ref{L-S-H} implies that
    \[d_S(U_j,W)\geq d_H(i(U_j),i(W))\geq 2 a_1 \geq 2 \delta.\]
    
    In conclusion, $\bigcup_{i=1}^4 \cC_i \cup  \tcC_2$ is an $(n, \sum_{i=1}^4|\cC_i|+|\tcC_2|, 2\delta, \{k\})_q$ CDC.
\end{IEEEproof}

\begin{theorem}\label{M-Insert2'}
    Let $\cC_1, \cC_2, \cC_3, $ and $ \hcC_4$ be the codes defined in Theorem \ref{Insert2'}, and let $\tcC_2$ be  the code defined in Theorem \ref{M-Insert2}. Then $\bigcup_{i=1}^3 \cC_i \cup \hcC_4 \cup \tcC_2$ is an $(n,\sum_{i=1}^3 |\cC_i| + |\hcC_4|+ |\tcC_2|,2\delta,\{k\})_q$ CDC.
\end{theorem}

\begin{IEEEproof}
    The demonstration is analogous to that of Theorem \ref{M-Insert2} and is thus omitted.
\end{IEEEproof}

To simplify the subsequent discussion, let  $\Delta(m,n,d)_q$ and  $\Delta(m,n,d;r)_q$  be equal to 1, when $\min\{m,n\}<d$. 

By applying Theorem \ref{Mix} and Theorem \ref{M-Insert}, we obtain the following result.

\begin{theorem}\label{T-Mul-Mix}
Let $n, n_1, n_2, k, \delta$ be integers with $n=n_1+n_2$, $n_1 \geq k$, $n_2 \geq k$, and $k \geq 2\delta \geq 4$. Let $T_1 \subseteq [\delta,k]$ with $l_{T_1} <2 \delta$. Let $T_2 \subseteq [k+\delta-T_1^{\min},k]$ with $l_{T_2} < 2\delta$. For $i = 1,2$, let $\cX_i$ be an $(n_i, 2\delta, 2\delta-l_{T_i}, T_i)_q$ MDDC. Suppose that $n_1+T_2^{\min}-k 
\geq k -\delta$. Set 
 $\theta=\lfloor \frac{n_1+T_2^{\min}-2k+2\delta}{\delta} \rfloor $, 
$\zeta=\lfloor \frac{n_2+T_1^{\min}-k}{\delta}\rfloor$, and $\mu=2k-T_1^{\min}-T_2^{\min}$. Let $\lambda_i=n_1+T_2^{\min}-2k+(2-i)\delta$ 
 for $1\leq i \leq \theta$,  and let $\gamma_j=n_2+T_1^{\min}-k-j\delta$ for $1 \leq j \leq\zeta$. Then
    \begin{align*}
        A_q(n,2\delta,\{k\}) &\geq 
        \sum_{t \in T_1}\eta_t(\cX_1) \cdot \Delta(k,n_2+t-k,\delta)_q \\& +\sum_{s \in T_2}\eta_s(\cX_2) \cdot \Delta(k,n_1+s-k,\delta; T_1^{\min}-\delta-(k-s))_q
        \\ &+\Delta(k-\delta,n_2-T_2^{\min}+k-\delta,\delta; k-2\delta)_q \cdot \sum_{i=1}^{\theta}\sum_{j=1}^{\zeta} \Delta(\lambda_i,k-\delta,\delta)_q \cdot \Delta(\gamma_j,\delta,\delta)_q.
    \end{align*}   
\end{theorem}
\begin{IEEEproof}
Let $\cC_1 \cup \cC_2$ be the same as that in Theorem \ref{Mix}. Let $\tcC_1$ be constructed using the multilevel construction stated in Theorem \ref{M-Insert} whose  identifying vectors set is 
\begin{align*}
    \cS =\{\bv_{i,j}=(\underbrace{\overbrace{0\cdots0}^{(i-1)\delta}  \overbrace{1\cdots1}^{k-\delta} 0\cdots0}_{n_1+T_2^{\min}-k} \underbrace{0\cdots0}_{\mu} \underbrace{\overbrace{0\cdots0}^{(j-1)\delta}\overbrace{1\cdots1}^{\delta}0\cdots0 }_{n_2+T_1^{\min}-k}): 1 \leq i \leq \theta, 1\leq j \leq \zeta\}.
\end{align*}
It is clear that the identifying vectors in $\cS$ satisfy the conditions in Theorem \ref{M-Insert}.

For $\bv_{i,j} \in \cS,$ the Ferrers diagram of $EF(\bv_{i,j})$ is \[ \cF_{i,j}=[\underbrace{k-\delta, \dots, k-\delta}_{\lambda_i},
\underbrace{
\overbrace{k-\delta, \dots, k-\delta}^{\mu+(j-1)\delta} \overbrace{k,\dots,k}^{\gamma_j}}_{n_2-T_2^{\min}+k-\delta} ].\]
By Theorem \ref{M-Insert}, we require the existence of an $(\cF_{i,j},\delta;k-\delta,n_2-T_2^{\min}+k-\delta,k-2\delta)_q$ RFDRMC $\cC_{\cF_{i,j}}$ for all $\bv_{i,j} \in \cS$. It is clear that
\[\cF_{i,j} = \begin{pmatrix}
    \cF_i^1 & \cF^2\\
    & \cF_j^3
\end{pmatrix},\]
where $\cF_i^1=[\underbrace{k-\delta,\dots,k-\delta}_{\lambda_i}]$, $\cF^2=[\underbrace{k-\delta,\dots,k-\delta}_{n_2-T_2^{\min}+k-\delta}]$,  and $\cF_j^3=[\underbrace{\delta,\dots,\delta}_{\gamma_j}]$. By Proposition \ref{P-RFDRMC}, we can construct an $(\cF_{i,j}, \rho_{i,j}, \delta;k-\delta,n_2-T_2^{\min}+k-\delta,k-2\delta)_q$ RFDRMC $\cC_{\cF_{i,j}}$ with
\[\rho_{i,j} =\Delta(\lambda_i,k-\delta,\delta)_q \cdot \Delta(k-\delta,n_2-T_2^{\min}+k-\delta,\delta; k-2\delta)_q \cdot  \Delta(\gamma_j,\delta,\delta)_q.\]

Then, we construct an $(n,2\delta,\{k\})_q$ CDC  $\tcC_1 = \bigcup_{i=1}^{\theta}\bigcup_{j=1}^{\zeta} \sL(\cC_{\cF_{i,j}})$ with cardinality $\sum_{i=1}^{\theta}\sum_{j=1}^{\zeta} \rho_{i,j}$. Applying Theorem \ref{M-Insert}, $\cC_1 \cup \cC_2 \cup \tcC_1$ is our desired CDC. 
\end{IEEEproof}

Utilizing Theorems \ref{Insert2} and \ref{M-Insert2}, we derive the following result.

\begin{theorem}\label{T-Mul-Intert2}
Let $n, n_1, n_2, k, \delta$ be integers with $n=n_1+n_2$, $n_1 \geq k$, $n_2 \geq k$, and $k \geq 2\delta \geq 4$. Let $T_1 \subseteq [\delta,k]$ with $l_{T_1} <2 \delta$. Let $T_2 \subseteq [k+\delta-T_1^{\min},k]$ with $l_{T_2} < 2\delta$. For $i = 1,2$, let $\cX_i$ be an $(n_i, 2\delta, 2\delta-l_{T_i}, T_i)_q$ MDDC. Let $t_1, t_2, a_1,a_2,b_1$, and $b_2$ be positive integers satisfying $ a_1+a_2=k,$ $ b_1+b_2 \geq \delta$,  $a_1 \leq t_1 \leq n_1-k+T_2^{\min}-\delta$, 
    $a_2+k-T_1^{\min} \leq t_2 \leq n_2- \delta$, $n_2-t_2 \geq a_2$, $k-t_1\geq \delta$, and $a_i \geq \delta$,  $1 \leq b_i \leq \delta$ for $i = 1, 2$.    Suppose that 
 $n_1+T_2^{\min}-k-t_1 \geq \delta$ and $n_2+T_1^{\min}-k \geq k-\delta$. Set $\theta=\lfloor \frac{n_1+T_2^{\min}-k-t_1}{\delta}\rfloor$,  $\zeta=\lfloor \frac{n_2+T_1^{\min}-2k+2\delta}{\delta}\rfloor $,  and $\mu = 2k-T_1^{\min}-T_2^{\min}$. Let $\lambda_i=n_1+T_2^{\min}-k-t_1-i\delta$ for $1 \leq i \leq \theta$, and let $\gamma_j=n_2+T_1^{\min}-2k+(2-j)\delta$ for $1 \leq j \leq \zeta$. Then
    \begin{align*}
        A_q(n, 2\delta,\{k\}) &\geq \sum_{t \in T_1}\eta_t(\cX_1) \cdot \Delta(k,n_2+t-k,\delta)_q \\& +\sum_{s \in T_2}\eta_s(\cX_2) \cdot \Delta(k,n_1+s-k,\delta; T_1^{\min}-\delta-(k-s))_q  \\&+ A_q(t_1,2\delta, \{a_1\})\cdot \Delta(a_1, n_2-k+T_1^{\min}-a_2, \delta;a_1-\delta )_q  \\&\cdot\Delta(a_2, n_1-k+T_2^{\min}-a_1, \delta; a_2-\delta)_q\cdot A_q(t_2-k+T_1^{\min},2\delta, \{a_2\}) \\& \cdot f \cdot \Delta(a_1, n_1 -k+T_2^{\min}-t_1, \delta)_q \cdot \Delta(a_2, n_2-t_2, \delta)_q\\
        & +A_q(t_1,2\delta,\{a_1\})_q \cdot A_q(n_2-t_2,2\delta,\{a_2\})
        \cdot 
        \Delta(a_2,n_1-k+T_2^{\min}-a_1,\delta;a_2-\delta)_q  \\
        &\cdot \Delta(a_1, n_1-k+T_2^{\min}-t_1, \delta)_q
        \cdot
      \Delta(a_1,n_2-k+T_1^{\min}-a_2,\delta;a_1-\delta)_q \\ &\cdot \Delta(a_2,t_2-k+T_1^{\min},\delta; k-t_1-\delta)_q +
    \sum_{i=1}^{\theta} \sum_{j=1}^{\zeta} \Delta(\lambda_i, \delta,\delta)_q \cdot \Delta(\gamma_j, k-\delta, \delta)_q,
    \end{align*}
    where $f=\min\left\{\frac{\Delta(a_1,n_1-k+T_2^{\min}-t_1,b_1)_q}{\Delta(a_1,n_1-k+T_2^{\min}-t_1,\delta)_q},
        \frac{\Delta(a_2,n_2-t_2,b_2)_q}{\Delta(a_2,n_2-t_2,\delta)_q}
        \right\}$.
\end{theorem}
\begin{IEEEproof}
Let $\cC_1, \cC_2, \cC_3,$ and $\cC_4$ be constructed using Theorem \ref{Insert2}. Let $\tcC_2$ be constructed by the multilevel construction stated in Theorem \ref{M-Insert2} whose identifying vectors set is  
\begin{align*}
        \cS &=\{\tbv_{i,j}=(\overbrace{0\cdots0}^{t_1}\underbrace{\overbrace{0\cdots0}^{(i-1)\delta}\overbrace{1\cdots1}^{\delta}0\cdots0}_{n_1+T_2^{\min}-k-t_1}\overbrace{0\cdots0}^{\mu}\underbrace{\overbrace{0\cdots0}^{(j-1)\delta}\overbrace{1\cdots1}^{k-\delta}0\cdots0}_{n_2+T_1^{\min}-k}): 1\leq i \leq \theta, 1\leq j \leq \zeta
        \}.
    \end{align*}
   It is easy to verify that the identifying vectors in $\cS$ satisfy the conditions in Theorem \ref{M-Insert2}.
    
   For $\tbv_{i,j} \in \cS$, the Ferrers diagram of $EF(\tbv_{i,j})$ is 
    \[\tcF_{i,j}=[\overbrace{\delta,\dots,\delta}^{\lambda_i},\underbrace{\delta,\dots,\delta, \overbrace{k, \dots,k}^{\gamma_j}}_{n_2-T_2^{\min}+\delta}].\]  By Theorem \ref{M-Insert2}, we need to construct an $(\tcF_{i,j}, \delta; \delta, n_2-T_2^{\min}+\delta,0)_q$ RFDRMC $\cC_{\cF_{i,j}}$ for all $\tbv_{i,j} \in \cS$. 
 By Proposition \ref{P-RFDRMC}, we can construct an $(\tcF_{i,j}, \rho_{i,j}, \delta;  \delta, n_2-T_2^{\min}+\delta,0)_q$ RFDRMC $\cC_{\tcF_{i,j}}$ with
 \[
 \rho_{i,j}=\Delta(\lambda_i, \delta,\delta)_q \cdot \Delta(\gamma_j, k- \delta, \delta)_q.
 \]
    
    Let $\tcC_2 = \bigcup_{i=1}^{\theta}\bigcup_{j=1}^{\zeta} \sL(\cC_{\tcF_{i,j}})$ with cardinality $\sum_{i=1}^\theta \sum_{j=1}^{\zeta}\rho_{i,j}$. By Theorem \ref{M-Insert2}, $\bigcup_{i=1}^4\cC_{i} \cup \tcC_2$ is our desired CDC.
\end{IEEEproof}

Based on Theorems \ref{Insert2'} and  \ref{M-Insert2'}, the following conclusion is reached.

\begin{theorem}\label{T-Multi-Insert2'}
Let $n, n_1, n_2, k, \delta$ be integers with $n=n_1+n_2$, $n_1 \geq k$, $n_2 \geq k$, and $k \geq 2\delta \geq 4$.
Let $T_1 \subseteq [\delta,k]$ with $l_{T_1} <2 \delta$. Let $T_2 \subseteq [k+\delta-T_1^{\min},k]$ with $l_{T_2} < 2\delta$. For $i = 1,2$, let $\cX_i$ be an $(n_i, 2\delta, 2\delta-l_{T_i}, T_i)_q$ MDDC.
 Let $t_1, t_2, a_1,a_2,b_1,b_2,c_1$, and $c_2$ be positive integers satisfying $ a_1+a_2=k,$ $ b_1+b_2 \geq \delta$, $c_1+c_2\geq \delta$,  $a_1 \leq t_1 \leq n_1-k+T_2^{\min}-\delta$, 
    $a_2+k-T_1^{\min} \leq t_2 \leq n_2- \delta$, $n_2-t_2 \geq a_2$, $k-t_1\geq 2\delta$, and $a_i \geq \delta$,  $1 \leq b_i \leq \delta$, $1 \leq c_i \leq \delta$ for $i = 1, 2$. Suppose that 
 $n_1+T_2^{\min}-k-t_1 \geq \delta$ and $n_2+T_1^{\min}-k \geq k-\delta$. 
   Set $\theta=\lfloor \frac{n_1+T_2^{\min}-k-t_1}{\delta}\rfloor$,  $\zeta=\lfloor \frac{n_2+T_1^{\min}-2k+2\delta}{\delta}\rfloor $, and  $\mu = 2k-T_1^{\min}-T_2^{\min}$. Let  $\lambda_i=n_1+T_2^{\min}-k-t_1-i\delta$ for $1 \leq i \leq \theta$, and let $\gamma_j=n_2+T_1^{\min}-2k+(2-j)\delta$ for $1 \leq j \leq \zeta$. 
    Then 
    \begin{align*}
 A_q(n,2\delta,\{k\}) &\geq \sum_{t \in T_1}\eta_t(\cX_1) \cdot \Delta(k,n_2+t-k,\delta)_q \\& +\sum_{s \in T_2}\eta_s(\cX_2) \cdot \Delta(k,n_1+s-k,\delta; T_1^{\min}-\delta-(k-s))_q  \\&+A_q(t_1,2\delta, \{a_1\})\cdot \Delta(a_1, n_2-k+T_1^{\min}-a_2, \delta;a_1-\delta )_q  \\&\cdot\Delta(a_2, n_1-k+T_2^{\min}-a_1, \delta; a_2-\delta)_q\cdot A_q(t_2-k+T_1^{\min},2\delta, \{a_2\}) \\&\cdot f\cdot  \Delta(a_1, n_1 -k+T_2^{\min}-t_1, \delta)_q \cdot \Delta(a_2, n_2-t_2, \delta)_q\\
        &+ A_q(t_1, 2\delta,\{a_1\})\cdot A_q(n_2-t_2,2\delta,\{a_2\}) \cdot \Delta(a_2 ,n_1-k+T_2^{\min}-a_1, \delta;a_2-\delta)_q  \\&\cdot
    \Delta(a_1 ,n_2-k+T_1^{\min}-a_2, \delta; a_1-\delta)_q \cdot \Delta(a_1, n_1-k+T_2^{\min}-t_1, \delta)_q\\
    &\cdot[l \cdot \Delta(a_2,t_2-k+T_1^{\min},\delta;k-t_1-\delta)_q-(l-1)] \\ &+\sum_{i=1}^{\theta} \sum_{j=1}^{\zeta} \Delta(\lambda_i, \delta,\delta)_q \cdot \Delta(\gamma_j, k- \delta, \delta)_q,
    \end{align*}
    where $f=\min\left\{\frac{\Delta(a_1,n_1-k+T_2^{\min}-t_1,b_1)_q}{\Delta(a_1,n_1-k+T_2^{\min}-t_1,\delta)_q},
        \frac{\Delta(a_2,n_2-t_2,b_2)_q}{\Delta(a_2,n_2-t_2,\delta)_q}
        \right\}$ and \\  $l=\min \left\{\frac{\Delta(a_1,n_1-k+T_2^{\min}-t_1, c_1)_q}{\Delta(a_1,n_1-k+T_2^{\min}-t_1, \delta)_q}, \frac{\Delta(a_2 , t_2-k+T_1^{\min}, c_2)_q}{\Delta(a_2 ,t_2-k+T_1^{\min}, \delta)_q}\right\}$.
\end{theorem}

\begin{IEEEproof}
Let $\cC_1, \cC_2, \cC_3,$ and $\hcC_4$ be constructed from Theorem \ref{Insert2'}. Let $\tcC_2$ be the same as in the proof of Theorem \ref{T-Mul-Intert2}.
The proof is similar to that of Theorem \ref{T-Mul-Intert2} and is thus omitted.
\end{IEEEproof}

\begin{example}
Take the same notations as in Example \ref{E-18-4-6}.
    Set $\theta=1$, $\zeta=5$, and $\mu=1$. Let $\lambda_1=1$, 
   and let  $\gamma_j=10-2j$ for $1 \leq j \leq 5$. By Theorem \ref{T-Multi-Insert2'}, we choose the identifying vectors set as follows.
    \begin{align*}
        \cS=\{&\tbv_{1,1}=(001100111100000000), \tbv_{1,2}=(001100001111000000), \\& \tbv_{1,3}=(001100000011110000), \tbv_{1,4}=(001100000000111100), \\&\tbv_{1,5}=(001100000000001111)\}.
        \end{align*}
        Then we obtain that
        \begin{align*}
        |\tcC_2| &= \sum_{j=1}^5 \Delta(1,2,2)_q  \Delta(10-2j,4,2)_q=q^{24}+q^{18}+q^{12}+q^4+1.
        \end{align*}
        For $q=2$, $|\tcC_2|=17043473$. Thus
        \begin{align*}
            A_2(18,4,\{6\})&\geq|\cC_1 \cup \cC_2| + |\cC_3| + |\hcC_4| + |\tcC_2|\\
            &=1321068515713406208+133406654464+60548048+17043473\\&=1321068649197652193,
        \end{align*}
        which is better than the previously best lower bound $1321068515713406208$ in \cite{MixDD}.
\end{example}

\section{New lower bounds on the size of  CDCs}

In this section, we introduce new lower bounds on the size of CDCs derived from the integration of the mixed dimension construction, the inserting construction, and the multilevel construction discussed in Section \uppercase\expandafter{\romannumeral3}. We compare our proposed constructions with other previously established constructions of CDCs.

\subsection{New Lower Bounds of CDCs}

The following corollaries are derived from Theorem \ref{T-Mul-Mix}.

\begin{corollary}\label{C-Mulmixuseless}
    Let $\delta \geq 2$ and $k \geq 3\delta$ be two integers. Then 
    \begin{align*}
    A_q(2k,2\delta,\{k\}) &\geq q^{k(k-\delta+1)} + \Delta(k,k,\delta;k-\delta)_q+\Delta(k,k-\delta,\delta;k-2\delta)_qA_q(k,2\delta,\{k-\delta\})\\&+\Delta(k-\delta,k,\delta;k-2\delta)_q\sum_{j=1}^{\lfloor\frac{k}{\delta}\rfloor} \Delta(k-j\delta,\delta,
    \delta)_q.
    \end{align*}
\end{corollary}
\begin{IEEEproof}
    Set $n_1=k$, $n_2=k$, $T_1=\{k\}$, and $T_2=\{k,k-\delta\}$ in Theorem \ref{T-Mul-Mix}. If we construct the $(k,2\delta,\delta,T_2)_q$ MDDC from Proposition \ref{P-ConstrMDDC}, the desired conclusion follows.
\end{IEEEproof}

\begin{corollary}\label{2-12}
    Let $h$ be a non-negative integer. Then
    \begin{align*}
        A_2(12+h,4,\{4\}) &\geq 4801 \cdot 2^{12+3h}+327 \cdot 2^{\max\{3+h,4\}(\min\{3+h,4\}-1)}+A_2(4+h,4,\{4\})\\& + \sum_{i=1}^4 \sum_{j=1}^{\lfloor \frac{h+3}{2} \rfloor} \Delta(8-2i,2,2)_q \Delta(3+h-2j,2,2)_q.
    \end{align*}
\end{corollary}
\begin{IEEEproof}
    Set $q=2$, $n_1=8$, $n_2=4+h$, $\delta=2$, $k=4$, $T_1=\{4,3\}$, and $T_2=\{4\}$ in Theorem \ref{T-Mul-Mix}. If we take the $(8,4,3,\{4,3\})_2$ MDDC from Example \ref{Alg}, the conclusion follows.
\end{IEEEproof}

\begin{corollary}\label{q-12}
   Let $\delta \geq 2$ and $h \geq 0$ be two integers. If $N_q(4\delta, \delta,2\delta) > 0$, then
   \begin{align*}
       A_q(6\delta+h, 2\delta, \{2\delta\}) &\geq          A_q(4\delta,2\delta,\{2\delta\})q^{(2\delta+h)(\delta+1)}\\&+N_q(4\delta,\delta,2\delta)q^{\max\{\delta+h+1, 2\delta\}(\min\{\delta+h+1, 2\delta\} - \delta+1)}+A_q(2\delta+h,2\delta,\{2\delta\})\\&+\sum_{i=1}^4\sum_{j=1}^{\lfloor \frac{h+\delta+1}{\delta} \rfloor} \Delta(4\delta-i\delta, \delta, \delta)_q\Delta(h+\delta+1-j\delta,\delta,\delta)_q.             
   \end{align*}
\end{corollary}
\begin{IEEEproof}
    Set $n_1=4\delta$, $n_2=2\delta+h$, $k=2\delta$, $T_1=\{2\delta, \delta+1\}$, and $T_2=\{2\delta\}$ in Theorem \ref{T-Mul-Mix}. If we construct the $(4\delta, 2\delta,\delta+1, T_1)_q$ MDDC from Theorem \ref{T-ConstrMDDC}, the conclusion follows.
\end{IEEEproof}

By applying Theorems \ref{T-Mul-Intert2} and 
\ref{T-ConstrMDDC},  the following results are obtained.

\begin{corollary}\label{NEW2}
    Set $n_1=5$, $n_2=10$, $\delta=2$, $k=5$, $T_1=\{5\}$, $T_2=\{5,4\}$, $a_1=t_1=2$, $a_2=3$, $t_2=7$, and $b_1=b_2=1$. We have
    \begin{align*}
    A_q(15,4,\{5\}) &\geq q^{40}+ A_q(10,4,\{5\})(q^{16} + q^{15} + 2q^{14} + q^{13} - 2q^{11} - 3q^{10} -  4q^9 - 2q^8 \\&+ q^6 +  3q^5 + 2q^4 + q^3)+ N_q(10,2,5)(q^9 + q^8 + 2q^7 + q^6 + q^5 - q^4 \\&- q^3 - 2q^2 - q) + A_q(7,4,\{3\})q^{10} + q^{14}+q^{10}+q^6+q^2+1. 
    \end{align*}
    Set $n_1=5$, $n_2=11$, $\delta=2$, $k=5$, $T_1=\{5\}$, $T_2=\{5,4\}$, $a_1=t_1=2$, $a_2=3$, $t_2=7$, and $b_1=b_2=1$. We have
    \begin{align*}
      A_q(16,4,\{5\}) &\geq 
        q^{44}+ A_q(11,4,\{5\})(q^{16} + q^{15} + 2q^{14} + q^{13} - 2q^{11} - 3q^{10} -  4q^9 - 2q^8 \\&+ q^6 +  3q^5 + 2q^4 + q^3)+ N_q(11,2,5)(q^9 + q^8 + 2q^7 + q^6 + q^5 - q^4 \\&- q^3 - 2q^2 - q) + A_q(7,4,\{3\})q^{12}+ q^{16}+q^{12} + q^8 + q^3 + q^2 + 1.
    \end{align*}
    Set $n_1=5$, $n_2=12$, $\delta=2$, $k=5$, $T_1=\{5\}$, $T_2=\{5,4\}$, $a_1=t_1=2$, $a_2=3$, $t_2=9$, and $b_1=b_2=1$. We have
    \begin{align*}
      A_q(17,4,\{5\}) &\geq 
        q^{48}+ A_q(12,4,\{5\})(q^{16} + q^{15} + 2q^{14} + q^{13} - 2q^{11} - 3q^{10} -  4q^9 - 2q^8 \\&+ q^6 +  3q^5 + 2q^4 + q^3)+ N_q(12,2,5)(q^9 + q^8 + 2q^7 + q^6 + q^5 - q^4 \\&- q^3 - 2q^2 - q)+A_q(9,4,\{3\})q^{10}+q^{18} + q^{14} + q^{10} + q^6 + q^2 +1.
    \end{align*}
    Set $n_1=5$, $n_2=13$, $\delta=2$, $k=5$, $T_1=\{5\}$, $T_2=\{5,4\}$, $a_1=t_1=2$, $a_2=3$, $t_2=9$,  and $b_1=b_2=1$. We have
    \begin{align*}
      A_q(18,4,\{5\}) &\geq 
        q^{52}+ A_q(13,4,\{5\})(q^{16} + q^{15} + 2q^{14} + q^{13} - 2q^{11} - 3q^{10} -  4q^9 - 2q^8 \\&+ q^6 +  3q^5 + 2q^4 + q^3)+ N_q(13,2,5)(q^9 + q^8 + 2q^7 + q^6 + q^5 - q^4 \\&- q^3 - 2q^2 - q)+A_q(9,4,\{3\})q^{12}+q^{20} + q^{16} + q^{12} + q^8 + q^3 + q^2 + 1.
    \end{align*}
    Set $n_1=5$, $n_2=14$, $\delta=2$, $k=5$, $T_1=\{5\}$, $T_2=\{5,4\}$, $a_1=t_1=2$, $a_2=3$, $t_2=11$,  and $b_1=b_2=1$. We have
    \begin{align*}
      A_q(19,4,\{5\}) &\geq 
        q^{56}+ A_q(14,4,\{5\})(q^{16} + q^{15} + 2q^{14} + q^{13} - 2q^{11} - 3q^{10} -  4q^9 - 2q^8 \\&+ q^6 +  3q^5 + 2q^4 + q^3)+ N_q(14,2,5)(q^9 + q^8 + 2q^7 + q^6 + q^5 - q^4 \\&- q^3 - 2q^2 - q)+ A_q(11,4,\{3\})q^{10}+ q^{22} + q^{18} + q^{14} + q^{10} + q^6 + q^2 + 1.
    \end{align*}
    \end{corollary}

     Theorem \ref{T-Multi-Insert2'} together with Theorem \ref{T-ConstrMDDC}  provides some new lower bounds on CDCs.
    
    \begin{corollary}\label{New3}
Set $n_1=6$, $n_2=12$, $\delta=2$, $k=6$, $T_1=\{6\}$, $T_2=\{6,5\}$, $a_1=t_1=2$, $a_2=4$, $t_2=8$, $b_1=b_2=1$, and $c_1=c_2=1$. Then
        \begin{align*}
        A_q(18,4,\{6\}) &\geq q^{60}+A_q(12,4,\{6\})(q^{26} + q^{25} + 2q^{24} + q^{23} + q^{22} - q^{21} - 3q^{20} - 4q^{19}\\& - 3q^{18} - 2q^{17} + 4q^{15} + 5q^{14} + 5q^{13} + 3q^{12} + q^{11} - q^{10} - 3q^9 - 3q^8  \\&-2q^7 - q^6)+N_q(12,2,6)(q^{18} + q^{17} + 2q^{16} + 2q^{15} + q^{14} - q^{13} - 2q^{12} \\& -4q^{11} - 4q^{10} - 3q^9 - q^8 + q^7 + 2q^6 + 3q^5 + 2q^4 + q^3)+A_q(8,4,\{4\}) q^{18}\\&\cdot(q^6 + q^5 + q^4 - q^2 - q)+2q^{24} + 2q^{23} + 4q^{22} + 4q^{21} + 3q^{20} - q^{18} - 3q^{17} \\&- 3q^{16} - 3q^{15} - 4q^{14} - 4q^{13} - 2q^{12} + 2q^{10} + 4q^9 + 3q^8 + 2q^7 - q^5 + 1.
        \end{align*}
    
        Set $n_1=6$, $n_2=13$, $\delta=2$, $k=6$, $T_1=\{6\}$, $T_2=\{6,5\}$, $a_1=t_1=2$, $a_2=4$, $t_2=8$, $b_1=b_2=1$,  and $c_1=c_2=1$. Then
        \begin{align*} 
        A_q(19,4,\{6\}) &\geq q^{65}+A_q(13,4,\{6\})(q^{26} + q^{25} + 2q^{24} + q^{23} + q^{22} - q^{21} - 3q^{20} - 4q^{19}  \\&-3q^{18} - 2q^{17} + 4q^{15} + 5q^{14} + 5q^{13} + 3q^{12} + q^{11} - q^{10} - 3q^9 - 3q^8  \\&-2q^7 - q^6)+N_q(13,2,6)(q^{18} + q^{17} + 2q^{16} + 2q^{15} + q^{14} - q^{13} - 2q^{12} \\& -4q^{11} - 4q^{10} - 3q^9 - q^8 + q^7 + 2q^6 + 3q^5 + 2q^4 + q^3)+A_q(8,4,\{4\})(q^{27} \\&+ q^{26} + q^{25} - q^{23} - q^{22})+q^{27} + q^{24} + 2q^{23} + 4q^{22} + 5q^{21} + 3q^{20} - 2q^{18} -3q^{17} \\&-   3q^{16} - 2q^{15} - 4q^{14} - 4q^{13} - 3q^{12} + 2q^{10} + 4q^9 + 4q^8 + 2q^7 - q^5 - q^4 + 1.
        \end{align*}
    \end{corollary}

We list some of our new lower bounds in Table \ref{NB1}.

\begin{table}[!htp]
  \centering
  \caption{New lower bounds of CDCs}
  \label{NB1}
  \begin{tabular}{|c|c|}
  \hline
   New lower bounds ($q=2,3,4,5,7,8,9$) & Corollary \\ \hline
 $A_2(12,4,\{4\})$ & 
Corollary \ref{2-12} \\ \hline
$A_q(12,4,\{4\})$ ($q \geq 3$),
$A_q(19,6,\{6\})$ & Corollary \ref{q-12} \\ \hline
$A_q(n,4,\{5\}) (15 \leq n \leq 19)$& Corollary \ref{NEW2} \\ \hline
$A_q(n,4,\{6\}) (n=18,19)$& Corollary \ref{New3} \\ \hline 
  \end{tabular}
\end{table}


\subsection{Comparison of Our Lower Bounds with \cite{MixDD}}

In the following corollary, we compare the cardinalities of our codes obtained by Corollary \ref{C-Mulmixuseless} with the cardinality of codes obtained by the mixed dimension construction in  \cite[Corollary 5.5]{MixDD}, and provide a lower bound for their difference.

\begin{corollary}
     Let $q \geq 2$ be any prime power and $\delta$, $k$ be two integers with $\delta \geq 2$ and $k \geq 3\delta$. Let $\cC^0$ be a $(2k,|\cC^0|,2\delta,\{ k \})_q$ CDC obtained by \cite[Corollary 5.5]{MixDD} and $\cC$ be a $(2k,|\cC|,2\delta,\{ k \})_q$ CDC obtained by Corollary \ref{C-Mulmixuseless}, then
     \begin{align*}
         |\cC|-|\cC^0| > q^{(\delta+1)k-2\delta^2-\delta}(q^k-1).
     \end{align*}
\end{corollary}
\begin{IEEEproof} By \cite[Lemma 4]{Network}, the Guassian coefficient $\left[\begin{smallmatrix}
    n\\ l
\end{smallmatrix}\right]_q$ satisfies $\left[\begin{smallmatrix}
    n \\ l
\end{smallmatrix}\right]_q > q^{l(n-l)}$.
Then it can be easily verified that
\begin{align*}
   |\cC|-|\cC^0| &= \Delta(k-\delta,k,\delta;k-2\delta)_q\sum_{j=1}^{\lfloor\frac{k}{\delta}\rfloor} \Delta(k-j\delta,\delta,
    \delta)_q\\
    &> D(k-\delta,k,\delta,\delta)_q \Delta(k-\delta,\delta,\delta)_q\\&
    = \left[\begin{smallmatrix}
    k-\delta \\ \delta
\end{smallmatrix}\right]_q (q^k-1)q^{k-\delta} > q^{(\delta+1)k-2\delta^2-\delta}(q^k-1).
    \end{align*}
\end{IEEEproof}

From Corollaries \ref{2-12}, \ref{q-12},  \ref{NEW2} and \ref{New3}, we obtain some new lower bounds for CDCs. We
 list the improved lower bounds for CDCs with $q=2, 3$ in Table \ref{NewBound}
 and compare them with the previously best known lower bounds in \cite{MixDD}.

\begin{table}[!htp]
  \centering
  \caption{Comparison of cardinalities of our codes with the codes in \cite{MixDD}}
  \label{NewBound}
  \begin{tabular}{|c|c|c|}
  \hline
   $A_q(n,d,k)$ & New & Old \\ \hline
$A_2(12,4,\{4\})$ & 19748694 & 19748609\\ \hline
 $A_2(15,4,\{5\})$ & 
 1252457773879 &1252457415410 \\ \hline
 $A_2(16,4,\{5\})$ & 
 20021970273665
 &20021968839796 \\ \hline
 $A_2(17,4,\{5\})$ & 320366594803351 &320366588394066 \\ \hline
 $A_2(18,4,\{5\})$ & 5125925812094397 & 5125925786457264\\ \hline
 $A_2(19,4,\{5\})$ & 82014832991711141&
 82014832887370592\\ \hline
 $A_2(18,4,\{6\})$ & 1321068649197652193 & 1321068515713406208\\ \hline
 $A_2(19,4,\{6\})$& 42242647556085589201 &
42242646488635457536\\ \hline
$A_2(19,6,\{6\})$&4527333206655562&4527333206654977\\ \hline
$A_3(12,4,\{4\})$&288652606436 & 288652605616\\
\hline
  \end{tabular}
\end{table}

\section{Conclusion}
Although numerous constructions have provided new lower bounds for CDCs, there still exists a significant gap between the lower bounds and the upper bounds on the size of CDCs. Initially, most constructions of CDCs rely on smaller CDCs and RMCs, where the inserting construction and the multilevel construction are very effective techniques. In this paper, we generalize the inserting construction and multilevel construction, and combine them with the mixed dimension construction which is based on smaller MDDCs and RMCs. When $k \geq 2\delta$, our constructions improve the mixed dimension construction and lead to some new lower bounds on the sizes of CDCs for various parameters. 
 
When the parameters are further restricted, our codes can accommodate additional codewords. There is a $k \times (2k -T_1^{\min}-T_2^{\min})$ zero matrix in the generator matrix of subspaces from our inserting constructions, but we don't take it into account when performing the multilevel construction.  Incorporating this zero matrix into the analysis may lead to even better outcomes.
This paper presents several methods for selecting qualified identifying vectors. However, determining an effective method for selecting identifying vectors is an open and challenging issue.

For further research,  we are interested in how to improve the mixed dimension construction when $k < 2\delta$. RFDRMCs and MDDCs play important roles in our constructions. In the future, one may focus on studying the bounds and constructions of RFDRMCs and MDDCs.

%

\end{document}